\documentclass[fleqn,usenatbib,usedcolumn]{mnras}

\usepackage[british]{babel}

\usepackage{ae,aecompl}
\usepackage{newtxtext}
\usepackage{newtxmath}
\usepackage[T1]{fontenc}

\usepackage{graphicx}
\usepackage{amsmath}

\newcommand{\bagpipes}{\textsc{Bagpipes}}
\newcommand{\mufasa}{\textsc{Mufasa}}

\title[Inferring the SFHs of massive quiescent galaxies]{Inferring the star-formation histories of massive quiescent galaxies with BAGPIPES: Evidence for multiple quenching mechanisms}

\author[A. C. Carnall et al.]{
A. C. Carnall,$^{1}$\thanks{E-mail: adamc@roe.ac.uk}
R. J. McLure,$^{1}$
J. S. Dunlop,$^{1}$
R. Dav\'{e},$^{1}$
\\
$^{1}$SUPA\thanks{Scottish Universities Physics Alliance}, Institute for Astronomy, University of Edinburgh, Royal Observatory, Edinburgh EH9 3HJ, UK
\\
}

\date{Accepted XXX. Received YYY; in original form ZZZ}

\pubyear{2017}

\begin{document}
\label{firstpage}
\pagerange{\pageref{firstpage}--\pageref{lastpage}}
\maketitle

\begin{abstract}
We present Bayesian Analysis of Galaxies for Physical Inference and Parameter EStimation, or \bagpipes, a new \textsc{Python} tool which can be used to rapidly generate complex model galaxy spectra and to fit these to arbitrary combinations of spectroscopic and photometric data using the \textsc{MultiNest} nested sampling algorithm. We extensively test our ability to recover realistic star-formation histories (SFHs) by fitting mock observations of quiescent galaxies from the \mufasa\ simulation. We then perform a detailed analysis of the SFHs of a sample of 9289 quiescent galaxies from UltraVISTA with stellar masses, $M_* > 10^{10}\ \mathrm{M_\odot}$ and redshifts $0.25 < z < 3.75$.  The majority of our sample exhibit SFHs which rise gradually then quench relatively rapidly, over $1{-}2$ Gyr. This behaviour is consistent with recent cosmological hydrodynamic simulations, where AGN-driven feedback in the low-accretion (jet) mode is the dominant quenching mechanism. At $z > 1$ we also find a class of objects with SFHs which rise and fall very rapidly, with quenching timescales of $< 1$ Gyr, consistent with quasar-mode AGN feedback. Finally, at $z < 1$ we find a population with SFHs which quench more slowly than they rise, over $>3$ Gyr, which we speculate to be the result of diminishing overall cosmic gas supply. We confirm the mass-accelerated evolution (downsizing) trend, and a trend towards more rapid quenching at higher stellar masses. However, our results suggest that the latter is a natural consequence of mass-accelerated evolution, rather than a change in quenching physics with stellar mass. We find $61\pm8$ per cent of $z > 1.5$ massive quenched galaxies undergo significant further evolution by $z = 0.5$. \bagpipes\ is available at \href{https://bagpipes.readthedocs.io}{bagpipes.readthedocs.io}.
\end{abstract}

\begin{keywords}
galaxies: evolution -- galaxies: star formation -- methods: statistical
\end{keywords}

\section{Introduction}\label{sect:introduction}

Recent years have seen rapid advances in the quantity and quality of data available for the study of galaxy evolution. Where once studies of large samples of objects relied solely on photometric data, the advent of multi-object and integral-field spectrographs, and the advancing near-infrared capability of both spectroscopic and photometric instruments mean that increasingly detailed studies can now be performed on statistical samples of objects. This presents new challenges, as ever more complex models must be developed to accurately reproduce observed properties, and more advanced statistical techniques are necessary to deal with larger sample sizes and the higher-dimensional, often highly degenerate parameter spaces of more complex models. Techniques from the field of data science for dealing with large datasets, such as machine learning, are also becoming increasingly valuable to astronomers.

One of the most striking observations of recent years is the bimodal distribution in rest-frame colours which galaxies have been observed to display across at least the $10-11$ billion years since redshift, $z = 2-3$ (e.g. \citealt{Strateva2001}; \citealt{Bell2004}; \citealt{Faber2007}; \citealt{Williams2009}; \citealt{Brammer2009}; \citealt{Whitaker2011, Whitaker2013}; \citealt{Straatman2014, Straatman2016}). When plotted on a colour-magnitude diagram, this bimodality manifests as a diffuse blue cloud and tight red sequence, with a sparsely populated green valley in between. In addition, the same bimodality has been observed in several other properties, such as environment (e.g. \citealt{Baldry2006}; \citealt{Hartley2010}; \citealt{Chuter2011}; \citealt{Law-Smith2017}) and morphology (e.g. \citealt{Papovich2012}; \citealt{Bell2012}; \citealt{Strazzullo2013}; \citealt{vanderwel2014}), with blue, star-forming galaxies exhibiting a more disk-like morphology and being less strongly clustered than red, quiescent galaxies, which typically exhibit early-type morphologies.

The observation of this bimodality has led to the search for a quenching mechanism, or mechanisms, capable of shutting down star-formation activity in galaxies. Many potential quenching mechanisms have been proposed, falling broadly into two categories (e.g. \citealt{Peng2015}). We refer to these as ejective, associated with the expulsion of the gas reservoir which fuels star formation, and preventative, associated with the supply of new gas to the galaxy being shut down. The former results in a rapid halting of star formation, whereas the latter leads to a more gradual decline, or strangulation, as the existing reservoir is depleted.

The most commonly discussed process related to the quenching of massive galaxies is active galactic nucleus (AGN) feedback, in which the accretion of gas onto a galaxy fuels an AGN. This, in turn, either ejects the gas reservoir from the galaxy, or prevents further gas from being accreted, depending on the strength of the feedback. Strong AGN-driven outflows, sometimes called quasar-mode feedback, have been observed at high redshifts (e.g. \citealt{Maiolino2012}; \citealt{forster-schreiber2014}), and are thought, in some cases, to be triggered by major-merger events. However, it has been shown that galaxies quenched in this way are likely to re-ignite star formation (e.g. \citealt{Gabor2010}), rather than remaining quiescent.

Sustained quiescence is thought to require a continuing input of energy to heat the circumgalactic medium, preventing new gas from falling onto the galaxy (e.g. \citealt{Gabor2011}). Low-accretion-state AGN feedback, sometimes called radio-mode or jet-mode, has been proposed \citep{Croton2006} as the source of this energy input. This mechanism has been shown to significantly improve agreement with the observed properties of the red sequence when implemented in modern hydrodynamic simulations, such as \mufasa\ \citep{Dave2017} and IllustrisTNG \citep{Nelson2017}.

Another example of an ejective quenching process is the tidal/ram-pressure stripping experienced by galaxies falling into clusters. This process is also capable of causing very rapid quenching of star formation, and is thought to to give rise to the environmental dependence of the galaxy colour bimodality (e.g. \citealt{Peng2012}). Rapid quenching of satellites is also observed in simulations such as Illustris (e.g. \citealt{Diemer2017}).

However, recently \cite{Abramson2016} have shown that no qualitatively different processes are required to have acted on different galaxies to explain the galaxy colour bimodality. In their model, quiescence is the end-point of a single evolutionary track along which all galaxies progress at different rates. Star formation in individual galaxies dies down naturally in the same way as the cosmic star formation rate (SFR) density. Under this paradigm, we should not necessarily expect to observe the imprints of quenching processes in the properties of quiescent galaxies.

In order to understand whether, and to what extent, different potential quenching processes contribute to the termination of star formation, much tighter observational constraints on the way quenching behaves will be necessary. Two fundamental questions which we will seek to address in this work are:
\begin{enumerate}
\item When did quenched galaxies form their stellar mass?
\item How long did the process of quenching take?
\end{enumerate}

\noindent It is already known that a complete answer to these questions must include a description of the dependencies on physical properties such as size and mass, and the effects of different environments. Differences in the properties of quenched galaxies observed at different redshifts may also provide useful insights. In this work we will focus on observed-redshift and stellar-mass dependencies for massive quenched galaxies ($M_* > 10^{10}\ \mathrm{M_\odot}$).

Question (i) has been studied by extracting the star formation histories (SFHs) of local galaxies from the fossil records imprinted in their spectra (e.g. \citealt{Heavens2004}; \citealt{Panter2007}; \citealt{Thomas2010}; \citealt{Carson2010}; \citealt{Citro2016}). More recently this analysis has been extended to higher redshifts (e.g. \citealt{Moresco2010}; \citealt{Onodera2012, Onodera2015}; \citealt{Jorgensen2013}; \citealt{Gallazzi2014}; \citealt{Choi2014}; \citealt{Lonoce2014}; \citealt{Fumagalli2016}; \citealt{Pacifici2016}; \citealt{Siudek2017}) to connect local galaxies with their precursors at earlier epochs, and build up a consistent picture of quenching across cosmic time.

Whilst individual results for the derived ages of quenched galaxies are difficult to compare due to differences in the definition of age, a coherent picture has been established, often referred to as downsizing or mass-accelerated evolution, with more massive galaxies typically forming their stellar masses at earlier epochs than their less massive counterparts. An average trend towards later formation for samples observed  at lower redshifts is also seen, indicating that the assembly of the red sequence is still ongoing, in agreement with the observed evolution of the galaxy stellar mass function (e.g. \citealt{Tomczak2014}).

Question (ii) however is less well understood. \cite{Barro2013} and \cite{Schawinski2014} present evidence for a scheme in which there are both fast and slow tracks towards quiescence, with different mechanisms driving quenching. Massive quiescent galaxies at earlier epochs (z $\gtrsim$ 1.5) which are denser systems, colloquially known as `red nuggets', quench rapidly through strong AGN-driven outflows, whereas a second population of less-dense star-forming galaxies quench more gradually as a result of strangulation processes. The denser, earlier-quenching systems are theorised to gradually expand through processes such as minor mergers to leave one population of relatively large quiescent galaxies by the present epoch (e.g. \citealt{McLure2013}).

The results of \cite{Abramson2016} however show that it is not necessary to assume a change in quenching mechanism to obtain early-time rapid and late-time slower quenching.  This property is reproduced by the model of \cite{Gladders2013}, which assumes no physics, as part of a more general relationship, in which galaxies cross the green valley between star-forming and quiescent states in $\simeq20$ per cent of the age of the Universe at the epoch of their quenching. A similar result is also obtained by \cite{Pacifici2016}, although they also argue that their results are inconsistent with a single quenching mechanism acting in all circumstances.

Finally, \cite{Peng2015} argue that a comparison of metallicities between local quiescent and star-forming galaxies strongly favours a slow strangulation of star formation over a timescale of $\simeq4$ Gyr. Clearly, a direct method which could be demonstrated to reliably constrain in detail the SFHs of individual quiescent galaxies would greatly assist in reconciling these disparate results.

\begin{table*}
  \caption{Mean 5$\sigma$ limiting magnitudes within 2$^{\prime \prime}$ apertures for the 1 deg$^2$ twelve band catalogue of \protect \cite{Mortlock2017}. }
\begin{tabular}{lccccccccccccc}
\hline
Region & Percentage of area & $u^{*}$ &$g'$ &$r'$& $i'$&  $z'$ &$z_{\mathrm{Subaru}}'$ & $Y$& $J$ &$H$ & $K_{s}$ &IRAC1 & IRAC2 \\
\hline
Deep strips & 40\% & 27.0 & 27.1 & 26.6 & 26.3 & 25.4 & 26.4 & 25.1 & 24.9 & 24.6 & 24.8 & 25.3 & 25.1 \\
Wide strips & 60\% & 27.0  & 27.1  & 26.6  & 26.3  & 25.4  & 26.4  & 24.7 & 24.4 & 24.1 & 23.9 & 25.3  & 25.1 \\
\hline
\end{tabular}
\label{table:depths}
\end{table*}

Methods employed to obtain this information rely on modelling the light emitted by galaxies as a function of wavelength in terms of the physical parameters of the system (e.g. \citealt{CidFernandes2005}; \citealt{Thomas2017}; \citealt{Wilkinson2017}). These models are then fitted to observational data, which may consist of spectroscopy and/or photometry (sometimes referred to as the spectral energy distribution; SED). Once the posterior distribution for model parameters has been obtained, nuisance parameters may be marginalised out to obtain constraints on the parameters of interest, such as the SFH. A common approach has been to fit spectroscopic data indirectly, either by data compression (e.g. \citealt{Heavens2000}), or by using pre-calibrated spectral indices (e.g. \citealt{Worthey1994}).

Recently, the development of new Bayesian statistical techniques, such as advanced Markov chain Monte Carlo (MCMC; e.g. \citealt{Goodman2010}; \citealt{Foreman-Mackey2013}) and nested sampling algorithms (\citealt{Skilling2006}, \citealt{Feroz2008}; \citealt{Feroz2009, Feroz2013}) has begun to enable the efficient exploration of higher-dimensional parameter spaces to obtain posterior distributions for the parameters of complex models.

A new generation of modern spectral modelling and fitting tools has been developed in order to exploit this e.g. \textsc{Beagle} \citep{Chevallard2016} and \textsc{Prospector} (\citealt{Leja2017}; Johnson et al. in prep). These codes allow on-the-fly generation and fitting of complex, self-consistent models to describe galaxies across continuous parameter spaces. Their models include emission and absorption processes due to the stellar population, ionized gas in H\,\textsc{ii} regions, diffuse dust in the interstellar medium (ISM) and neutral gas in the intergalactic medium (IGM) along our line-of-sight. The ability to explore higher-dimensional parameter spaces also opens up the ability to fit more complex SFHs. This is important because it has been shown that the rigid, exponentially declining SFHs typically employed in SED fitting techniques can introduce significant biases into SFH parameter estimates (see Section \ref{sect:mufasa}).

The combination of new statistical techniques with the ever increasing volume and quality of data available means that it is now possible to study the behaviour of quenching processes in unprecedented detail. In this work we analyse photometric data for a large sample of quenched galaxies from the UltraVISTA Survey \citep{McCracken2012} using Bayesian Analysis of Galaxies for Physical Inference and Parameter EStimation, or \bagpipes, a new public galaxy spectral modelling framework and fitting tool written in the \textsc{Python} programming language. \bagpipes\ provides a highly intuitive application programming interface (API) for rapid, on-the-fly generation (up to hundreds of models per second) of complex, physically realistic model galaxy spectra across continuous parameter spaces. It also provides a tool, built around the \textsc{MultiNest} nested sampling algorithm (\citealt{Feroz2008}; \citealt{Feroz2009, Feroz2013}), which allows for the direct fitting of these models to arbitrary combinations of spectroscopic and photometric data.

We analyse our recovered SFHs for these objects in the context of Questions (i) and (ii) posed above in order to understand the quenching properties of our sample as a function of stellar mass and observed redshift across the majority of cosmic time. In a follow-up paper we will extend our analysis using \bagpipes\ to the direct fitting of spectroscopic data from VANDELS (\citealt{McLure2018}; \citealt{Pentericci2018}) to obtain better constraints.

The structure of this paper is as follows. In Section \ref{sect:uvista_data} we introduce the UltraVISTA dataset we will use to explore quenching. In Section \ref{sect:bagpipes} we describe the model generation and fitting methodologies employed by the \bagpipes\ code. In Section \ref{sect:mufasa} we consider different models within \bagpipes, in particular different SFH parameterisations, and test their ability to recover realistic SFHs by fitting mock observations of simulated quenched galaxies from the \mufasa\ suite of cosmological hydrodynamic simulations \citep{Dave2016}. By this process we define a model which is capable of recovering unbiased estimates of the SFH properties of large samples of quenched galaxies from photometric data, making use of a double-power-law SFH parameterisation. In Section \ref{sect:uvista_sample}, we apply this \bagpipes\ model to select a sample of 9289 quenched galaxies with $M_* > 10^{10}\ \mathrm{M_\odot}$ and redshifts in the range $0.25 < z < 3.75$ from UltraVISTA. In Section \ref{sect:results} we analyse the properties of their SFHs to constrain the epoch and duration of their quenching as a function of stellar mass and observed redshift. We present our conclusions in Section \ref{sect:conclusions}.

All magnitudes are quoted in the AB system and for all cosmological calculations we adopt $\Omega_M = 0.3$, $\Omega_\Lambda = 0.7$ and $H_0$ = 70 $\mathrm{km\ s^{-1}\ Mpc^{-1}}$. We distinguish between times, $t$, which are measured forwards from the beginning of the Universe (i.e. $t(z)$ is the age of the Universe at redshift $z$), and ages, $a$, which are measured backwards in time from the redshift of observation, $t(z_\mathrm{obs})$.

\section{The UltraVISTA Data}\label{sect:uvista_data}

The UltraVISTA Survey \citep{McCracken2012} is an ultra-deep imaging program over a 1.5 deg$^2$ contiguous area of the Cosmological Evolution Survey (COSMOS) field in the near-infrared $YJHK_s$ bands on the Visible and Infrared Survey Telescope for Astronomy (VISTA). The area is divided into deeper and shallower stripes (UltraVISTA deep and UltraVISTA wide, respectively) which each account for half of the total area. The data utilised here comes from the UltraVISTA DR3 release.

In this work we make use of the $K$-band selected catalogue compiled by \cite{Mortlock2017}, which covers the 1 deg$^2$ overlap region between UltraVISTA and the Canada-France-Hawaii Telescope Legacy Survey (CFHTLS; \citealt{Hudelot2012}) T0007 release, which provides optical $u^*g'r'i'z'$ band imaging.  These datasets are also combined with deep $z'$-band imaging from the Subaru telescope \citep{Furusawa2016} and 3.6 $\mathrm{\mu m}$ + 4.5 $\mathrm{\mu m}$ imaging from {\it Spitzer}/IRAC (\citealt{Ashby2013}; \citealt{Steinhardt2014}) to produce a final twelve-band catalogue. The process by which the catalogue is compiled is described in detail in sections 2.1, 2.3 and 2.5 of \cite{Mortlock2017}. Table \ref{table:depths} gives the mean depths for each of the twelve bands across the wide and deep regions, which respectively make up 60\% and 40\% of the 1 deg$^2$ overlap region.

\cite{Mortlock2017} also calculate highly robust photometric redshifts for their whole catalogue using the median values from five different photometric-redshift codes. These median photometric redshifts have $\sigma_{dz} \simeq 0.02$, where $dz = (z_{spec} - z_{phot})/(1 + z_{spec})$, and a catastrophic outlier fraction ($|dz| > 0.15$) of $\sim1\%$ (see their section 3 for more details). They also calculate mass-completeness limits as a function of redshift and clean the catalogue by performing star-galaxy separation, and by matching to radio and X-ray datasets to remove AGN contaminants. The combination of all of these high-quality datasets provides us with an extremely clean catalogue of unrivalled scope and depth for studying galaxy evolution.

\begin{table*}
  \caption{List of input parameters for the example \bagpipes\ model shown in Fig. \protect \ref{fig:example_pipes_model}. All parameters are discussed in Section \protect \ref{sect:bagpipes} except for $\alpha$, $\beta$ and $\tau$ for the double-power-law SFH parameterisation, the functional form for which is given in Equation \protect \ref{dblplaw}. Here we assume Solar metallicity to take the value $\mathrm{Z_\odot} = 0.02$.}
\begin{tabular}{ccccccc}
\hline
Global & & Double-power-law& & Dust & & Nebular\\

\hline
$\sigma_\mathrm{vel}$ = 300 km s$^{-1}$ & & log$_{10}\big(M_\mathrm{formed}/\mathrm{M_\odot}\big) = 11$  & & Calzetti & & log$_{10}(U) = -3$\\

$a_\mathrm{BC} = 0.01$ Gyr & & $Z= 0.8$ $Z_\odot$ & & $A_V = 0.2$ & &  \\

$z = 0$ & & $\tau = 12$ Gyr & & $\epsilon = 3$ & & \\

& & $\beta = 0.5$ & & $T = 30$ K & & \\

& & $\alpha = 30$ & & $\beta = 1.5$ & & \\

\hline
\end{tabular}
\label{table:example_model_params}
\end{table*}

\begin{figure*}
	\includegraphics[width=\textwidth]{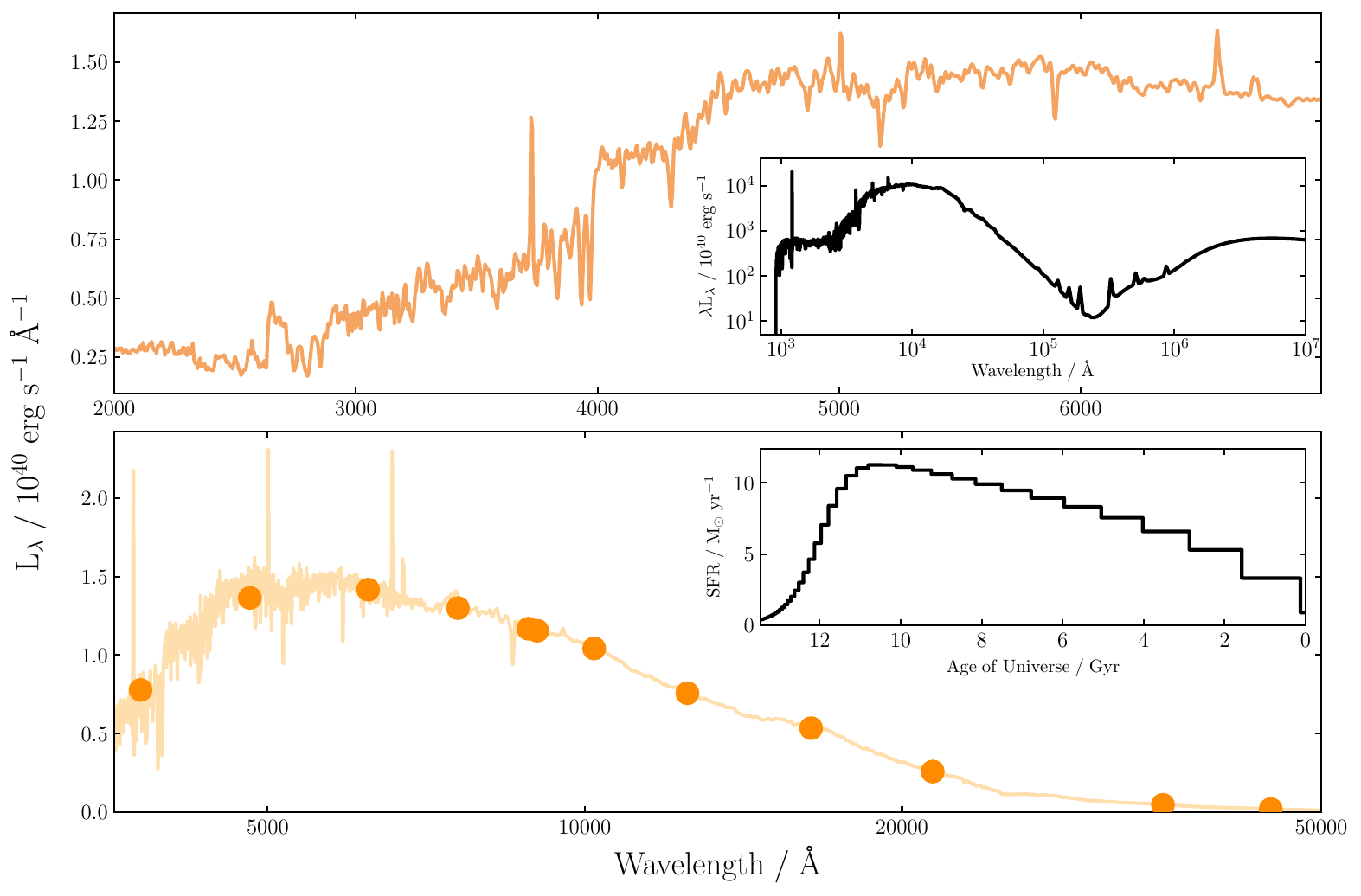}
    \caption{An example of a model galaxy built with \bagpipes. The parameters of the model are given in Table \protect \ref{table:example_model_params} and are designed to be representative of a present-day quenched galaxy. The top panel shows the spectral output, in this case over the range $2000{-}7000$ \AA\ with a sampling of 5\ \AA. The top inset panel shows the full spectrum for the model. The bottom panel shows the photometric output, in this case fluxes observed through the twelve filters of the UltraVISTA catalogue (orange circles) described in Section \ref{sect:uvista_data}. The bottom inset panel shows the SFH of the galaxy, which has been modelled as a double-power-law.}\label{fig:example_pipes_model}
\end{figure*}

\section{The BAGPIPES Code}\label{sect:bagpipes}

Bayesian Analysis of Galaxies for Physical Inference and Parameter EStimation, or \bagpipes, is a Bayesian spectral fitting code, designed to model the emission from galaxies from the far-ultraviolet to the microwave regimes, and to fit these models to arbitrary combinations of spectroscopic and photometric observational data using the \textsc{MultiNest} nested sampling algorithm (\citealt{Feroz2008}; \citealt{Feroz2009, Feroz2013}). \bagpipes\ is written purely in the \textsc{Python} programming language and considerable effort has been made to make the API as intuitive and user-friendly as possible. This section describes the \bagpipes\ code. In Section \ref{subsect:pipes_model} we describe how models are generated by the code, then in Section \ref{subsect:pipes_fitting} we describe the fitting of these models to observational data. \bagpipes\ is publicly available and fully documented at \href{https://bagpipes.readthedocs.io}{https://bagpipes.readthedocs.io}.

\subsection{Model generation}\label{subsect:pipes_model}

The first major aspect of \bagpipes\ is its ability to rapidly generate physically realistic model galaxy spectra from the far-ultraviolet to the microwave regime. Models can be built up to the desired level of complexity by specifying a number of model components, for example dust and nebular-emission prescriptions and SFH components. \bagpipes\ uses these components (in practice these are passed to the API as \textsc{Python} dictionaries containing parameter values) to generate an internal model of a galaxy spectrum. From this model, the user can request spectral data covering a given wavelength range at a given sampling, photometric fluxes through a series of user-defined filters, and emission-line fluxes. An example model is shown in Fig. \ref{fig:example_pipes_model}. The parameters of this model are listed in Table \ref{table:example_model_params}.

A model galaxy spectrum in \bagpipes\ observed at redshift $z_\mathrm{obs}$ has a luminosity per unit rest-frame wavelength, $\lambda$ of $L_\lambda (\lambda)$. The luminosity is constructed as a sum involving four ingredients:

\begin{enumerate}

\item Simple stellar-population models, $\mathrm{SSP}(a, \lambda, Z)$, which are a function of $\lambda$, the age of the stellar population, $a$, stellar metallicity, $Z$ and the initial mass function (described in Section \ref{subsubsect:pipes_sps}).

  \vspace{3mm}

\item The star-formation history, $\mathrm{SFR}(t)$, which is composed of a sum over one or more SFH components (described in Section \ref{subsubsect:pipes_sfh}).

  \vspace{3mm}

\item The transmission function of the ionized ISM, $T^+(a, \lambda)$, as defined by \cite{Charlot2001}, including absorption, line emission, ionized continuum emission and emission from warm dust within H\,\textsc{ii} regions (described in Section \ref{subsubsect:pipes_nebular}).

  \vspace{3mm}

\item The transmission function of the neutral ISM,  $T^0(a, \lambda)$, due to diffuse dust attenuation and emission (described in Section \ref{subsubsect:pipes_dust}).

\end{enumerate}

\noindent These four ingredients are described in more depth in the following four sections. The ingredients are then combined to give the galaxy luminosity according to the relationship

\begin{equation}\label{pipes_model_equation}
L_\lambda (\lambda) = \sum_{j=1}^{N_{c}} \sum_{i=1}^{N_a} \mathrm{SFR}_j(t_i)\ \mathrm{SSP}(a_i, \lambda, Z_j)\ T^+(a_i, \lambda)\ T^0(a_i, \lambda)\ \Delta a_i
\end{equation}

\noindent where $i$ runs across the age bins used in \bagpipes\ (see Section \ref{subsubsect:pipes_sps}), $\Delta a_i$ are the widths of these bins, $j$ runs across SFH components (see Section \ref{subsubsect:pipes_sfh}), $N_c$ is the number of SFH components and $N_a$ the number of age bins. The distinction between times and ages is explained at the end of Section \ref{sect:introduction}; here $t_i = t(z_\mathrm{obs}) - a_i$.

Once the galaxy luminosity has been calculated, it is redshifted and converted into an observed frame flux density, $f_{\lambda_\mathrm{obs}} (\lambda_\mathrm{obs})$, where $\lambda_\mathrm{obs} = (1+z_\mathrm{obs})\lambda$, using the relationship

\begin{equation}\label{eqn:redshift_spec}
f_{\lambda_\mathrm{obs}} (\lambda_\mathrm{obs}) = \frac{L_\lambda (\lambda)}{4\pi D_L(z_\mathrm{obs})^2 (1+z_\mathrm{obs})}T_{\mathrm{IGM}}(\lambda, z_\mathrm{obs}),
\end{equation}

\noindent where $D_L(z_\mathrm{obs})$ is the luminosity distance to redshift $z_\mathrm{obs}$ and $T_{\mathrm{IGM}}(\lambda, z_\mathrm{obs})$ is the transmission function of the IGM, described in Section \ref{subsubsect:pipes_igm}. Finally, if an output spectrum is requested (as opposed to photometry or emission-line fluxes), $L_\lambda (\lambda)$ can be convolved with a Gaussian kernel (before applying Equation \ref{eqn:redshift_spec}) to model the effects of velocity dispersion, as described in Section \ref{subsubsect:pipes_veldisp}.

\subsubsection{Stellar population synthesis}\label{subsubsect:pipes_sps}

Stellar Population Synthesis (SPS) is not implemented directly in \bagpipes. Instead, the code is designed to accept pre-defined SPS models in the form of grids of simple stellar-population (SSP) models of different ages across a range of metallicities.

The SPS models currently implemented in the code are the 2016 version of the \cite{Bruzual2003} (hereafter BC03) models. These differ from earlier versions by their use of the Medium-resolution Isaac Newton Telescope library of empirical spectra (MILES; \citealt{Falcon-Barroso2011}) in the UV-optical spectral region. The models implemented within the code are constructed using a \cite{Kroupa2001} initial mass function (IMF). \bagpipes\ does not currently include the option to vary element abundance patterns (e.g. alpha enhancement), with the model set currently implemented having scaled Solar abundances. In the future we also hope to distribute the code with the Binary Population and Stellar Synthesis (BPASS, \citealt{Eldridge2009}) models.

When a set of SPS models is loaded by \bagpipes, the models are resampled in age (using a weighted summation method) onto a grid of ages, $a_i$, from $\mathrm{log_{10}}(a_i/\mathrm{Gyr}) =$ 6.0 to 10.2 with uniform width in $\mathrm{log_{10}}(\Delta a_i/\mathrm{Gyr})$. The default spacing is 0.1 dex, which sets the value of $N_a$ in Equation \ref{pipes_model_equation} to 43.

\subsubsection{Star-formation histories}\label{subsubsect:pipes_sfh}

Star-formation histories in \bagpipes\ are constructed from one or more components, $j$, each of which specifies some functional form for star-formation rate as a function of time, $\mathrm{SFR}_j(t)$. The total SFR is given by summing over these components, i.e.

\begin{equation}
\mathrm{SFR}(t) = \sum^{N_c}_{j=1}\mathrm{SFR}_j(t).
\end{equation}

\noindent When $L_\lambda (\lambda)$ is calculated using Equation \ref{pipes_model_equation}, the $\mathrm{SFR}_j(t)$ are evaluated for all $t_i = t(z_\mathrm{obs}) - a_i$ where $a_i$ is less than $t(z_\mathrm{obs})$, i.e. the age of the stellar population is less than the age of the Universe at $z_\mathrm{obs}$.  We set $\mathrm{SFR}_j(t_i)$ to zero for all $t_i$ corresponding to ages greater than the age of the Universe.

One may specify an unlimited number of SFH components, each with an individual functional form (see Section \ref{subsect:mufasa_sfh_param} for further discussion). The options currently implemented are:

\begin{itemize}
\item Delta function
\item Constant
\item Exponentially declining (Equation \ref{tau})
\item Delayed exponentially declining
\item Log-normal
\item Double-power-law (Equation \ref{dblplaw})
\item Custom (directly input an array of SFR values).
\end{itemize}

\noindent In this way, \bagpipes\ can be used to generate a huge parameter space of possible SFHs, encompassing, for example, the non-parametric SFHs used by \cite{Leja2017} by the use of multiple constant components, or SFHs drawn from simulations of galaxy formation as in \cite{Pacifici2016}. SFHs from simulations may be inputted either by assigning a burst component to each star particle (as in Section \ref{subsect:mufasa_mock}), or by loading tabulated SFHs from simulations directly into the code as custom SFH components.

Each component also requires the specification of a total mass of stars formed, $M_\mathrm{formed}$, by that component over its whole history, and a metallicity value, $Z_j$, which is generated by linearly interpolating the  $\mathrm{SSP}(a_i, \lambda, Z_j)$ between different grids of models. Thus, metallicity evolution can be modelled by specifying different metallicities for multiple SFH components covering different epochs of cosmic time. \bagpipes\ also includes basic functionality for specifying a distribution function for the metallicities of stars in a galaxy, as in \cite{Leja2017}, as an alternative to linear interpolation between grids. In the future we will extend this scheme to allow full chemical-evolution histories to be specified, permitting metallicity distribution functions which are also a function of cosmic time.

\begin{figure}
	\includegraphics[width=\columnwidth]{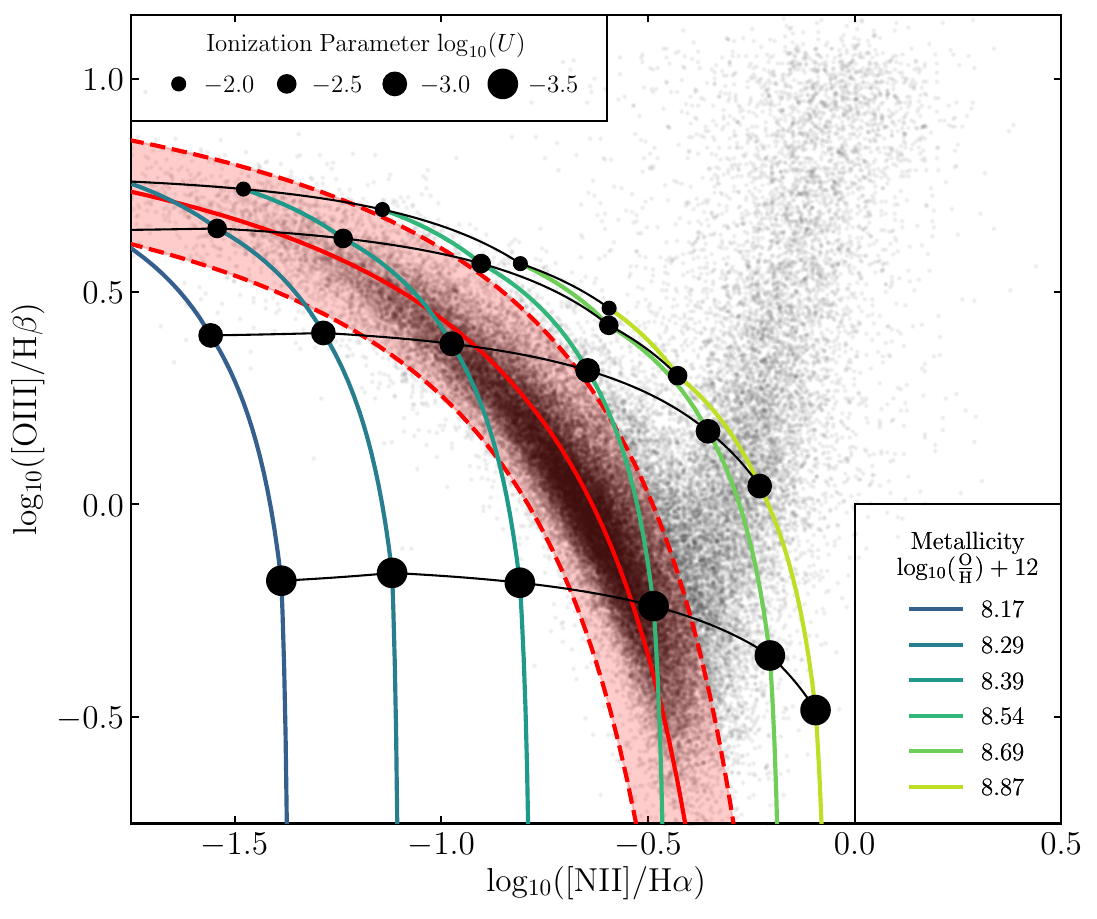}
    \caption{BPT diagram with the positions of \bagpipes\ models overlaid as a function of metallicity and ionization parameter. Black lines are lines of constant ionization parameter and coloured lines are lines of constant metallicity. The theoretically calibrated star-forming sequence for galaxies at $z\sim0$ of \protect \cite{Kewley2013} is shown in red and a sample of local SDSS galaxies from \protect \cite{Brinchmann2013} is shown in grey. The \bagpipes\ models are BC03-based, with constant SFR over the last 10 Myr. Our models show good overall agreement with the models of \protect \cite{Kewley2013} and the results of \protect \cite{Moustakas2006}, who find local star-forming galaxies fall within the range $-3.8$ \protect \textless $\mathrm{log_{10}}(U)$ \protect \textless $-2.9$ and $8.15 < \mathrm{12 + log\big(\frac{O}{H}\big)} < 8.7$.}\label{fig:bpt_figure}

\end{figure}

\subsubsection{Nebular emission}\label{subsubsect:pipes_nebular}

The nebular-emission model implemented in \bagpipes\ is constructed following the methodology of \cite{Byler2017}, using the latest (2017) version of the \textsc{Cloudy} photoionization code \citep{Ferland2017}. We model H\,\textsc{ii} regions with a spherical shell geometry of fixed radius and assume the nebular emission from a galaxy is the sum of emission from H\,\textsc{ii} regions of different ages, as in \cite{Charlot2001}. We assume the metallicity of the ionized gas is the same as the stars which produce the ionizing photons and use the Solar abundances of \cite{Anders1989} and ISM depletion factors and Helium and Nitrogen scaling relations of \cite{Dopita2000}. All of our models include dust grains using the ``ISM'' prescription within \textsc{Cloudy}, which has a grain size distribution and abundance pattern designed to reproduce the observed extinction properties for the ISM of the Milky Way.

\bagpipes\ includes functionality for computing grids of nebular-emission models corresponding to grids of input SSP models (these have been pre-computed for BC03). \textsc{Cloudy} is run using each SSP in turn as the input spectrum whilst varying the logarithm of the ionization parameter, $U$, in steps of 0.5 between $\mathrm{log}_{10}(U)$ = $-4$ and $-2$ by varying the number of hydrogen-ionizing photons, $Q_\mathrm{H}$ at a fixed Hydrogen density of 100 atoms cm$^{-3}$. At each value of $\mathrm{log}_{10}(U)$, $a_i$ and metallicity we save the output diffuse continuum, which includes contributions from ionized gas and warm dust, and fluxes for a series of emission lines.

The base list of lines we employ is that given in table 3 of \cite{Byler2017}, containing 128 separate features. The features have been renamed in the new version of \textsc{Cloudy}, due to slight shifts in wavelength, so it was necessary to manually match-up new and old labels. In addition, the five narrowly spaced C\,\textsc{ii} lines around 2326 \AA\ were replaced with the total flux for a blend of these lines, as identified by `Blnd 2326.00A', and the two narrowly spaced He\,\textsc{i} lines around 1.083 $\mathrm{\mu m}$ were replaced with `TOTL 1.08303m'. Finally we added the He\,\textsc{ii} feature at 4686 \AA, denoted in \textsc{Cloudy} by `He 2 4685.64A' to our list, meaning that, in total, we track 124 separate emission features.

When including nebular emission in a \bagpipes\ model, one must specify the ionization parameter and lifetime of H\,\textsc{ii} regions (or stellar birth-clouds), $a_\mathrm{BC}$. The nebular-continuum and line models corresponding to all $\mathrm{SSP}(a_i, \lambda, Z_j)$ with  $a_i < a_\mathrm{BC}$ are then linearly interpolated on-the-fly in $\mathrm{log}_{10}(U)$ and metallicity, and added to the corresponding $\mathrm{SSP}(a_i, \lambda, Z_j)$ model. Contrary to Equation \ref{pipes_model_equation} this is done by addition to, rather than multiplication of $\mathrm{SSP}(a_i, \lambda, Z_j)$, however the effect is the same as setting $T^+(a_i, \lambda)$ > 1.

\begin{figure}
	\includegraphics[width=\columnwidth]{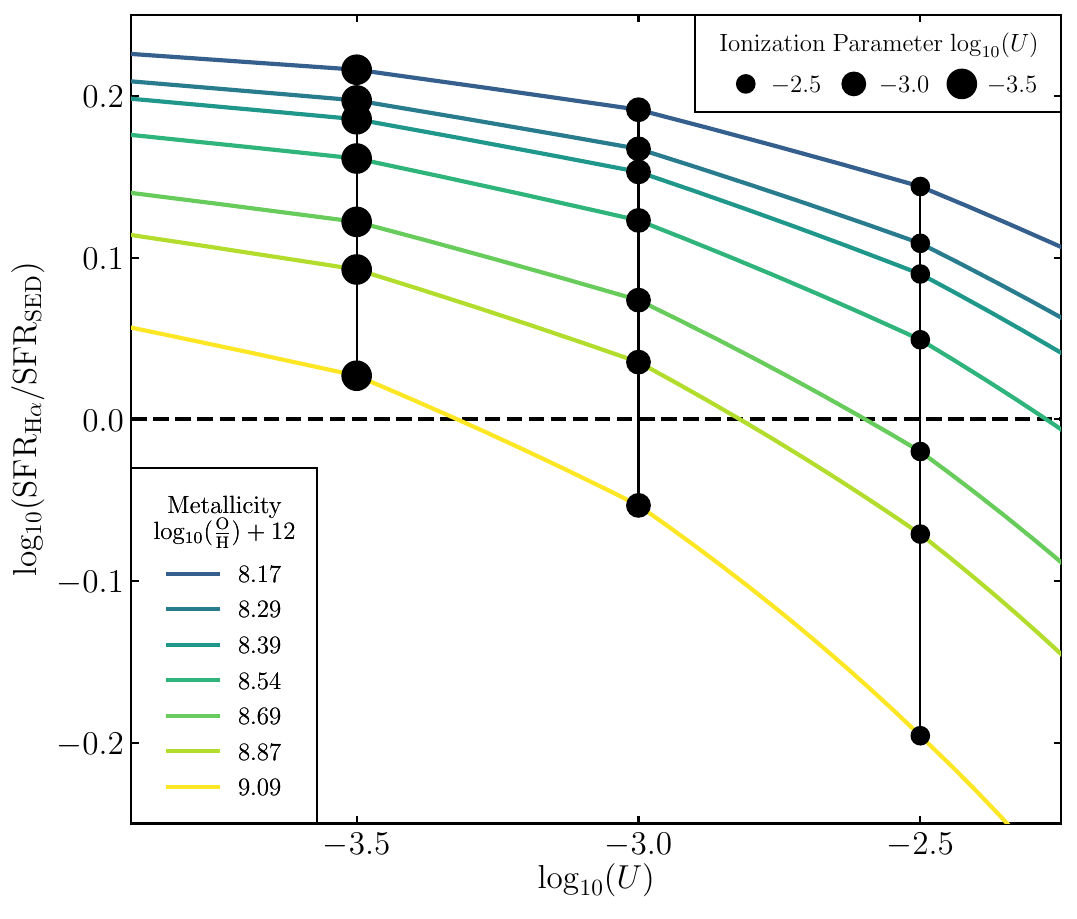}
    \caption{The offset between $\mathrm{H\alpha}$-derived SFR for \bagpipes\ models using the \protect \cite{Kennicutt2012} calibration and input SFR as a function of metallicity and ionization parameter. We find a slight excess of $\mathrm{H\alpha}$ flux in our models compared to those predicted by \protect \cite{Kennicutt2012}, except at supersolar metallicities or log-ionization parameters above $-2.5$. This is likely due to our assumption of an escape fraction of zero. Our models are consistent to within the observed scatter on the \protect \cite{Kennicutt2012} relationship.}\label{fig:kennicutt_figure}

\end{figure}

At this stage, the total energy in the combined (emission lines plus continuum) nebular model is set to the same as the total energy in the hydrogen-ionizing photons from all SSPs with $a_i < a_\mathrm{BC}$. The hydrogen-ionizing continuum is then removed from these SSPs (we set $T^+(a_i, \lambda)$ = 0 for $a_i < a_\mathrm{BC}$ and $\lambda < 911.8$ \AA; effectively assuming an escape fraction of zero) in order to maintain energy conservation. As well as being inserted into the output spectrum, observed fluxes for each emission feature are propagated through \bagpipes\ separately, meaning line fluxes can be accessed directly by the user.

We perform two consistency checks on our grid of BC03-based nebular-emission models to demonstrate that they are in agreement with similar predictions from the literature. We firstly check the optical line ratios as a function of metallicity and ionization parameter by plotting our models on the Baldwin-Phillips-Terlevich (BPT) diagram \citep{Baldwin1981} in Fig. \ref{fig:bpt_figure}. We confirm our predictions are consistent with those of \cite{Kewley2013} by comparison with their figure 1. We also demonstrate that the range of metallicities and ionization parameters measured by \cite{Moustakas2006} for local galaxies would be correctly reproduced by our models.

Having confirmed this, we then check that the line strengths are realistic in Fig. \ref{fig:kennicutt_figure} by comparing our input star-formation rate, SFR$_\mathrm{SED}$ with those derived from the strengths of our $\mathrm{H\alpha}$ lines using the calibration of \cite{Kennicutt2012}, SFR$_\mathrm{H\alpha}$. We find that our $\mathrm{H\alpha}$  fluxes tend to be slightly higher than those predicted by \cite{Kennicutt2012}, e.g. by 0.08 to 0.19 dex for metallicities of $\mathrm{12 + log\big(\frac{O}{H}\big)} = 8.69$ and 8.17 respectively at $\mathrm{log}_{10}(U) = -3.0$, which is consistent to within the observed scatter on the relationship. This overestimation is likely due to our assumption of an escape fraction of zero; an escape fraction of $\sim20$ per cent would be required to bring our $\mathrm{12 + log\big(\frac{O}{H}\big)} = 8.69$, $\mathrm{log}_{10}(U) = -3.0$ result into line with \cite{Kennicutt2012}.

\subsubsection{Dust attenuation and emission}\label{subsubsect:pipes_dust}

Dust attenuation in \bagpipes\ is designed in a modular fashion such that different dust curves may be easily implemented. Each dust-attenuation curve takes the overall functional form

\begin{equation}
\quad \mathrm{log_{10}}\Big(T^0(a_i, \lambda)\Big)\ =\ \begin{cases}
    \ \ \ -\frac{0.4\ \epsilon\ A_V\ k(\lambda)}{R_V} & \quad  a_i < a_\mathrm{BC}\\
    \ \ \ \ \ -\frac{0.4\ A_V\  k(\lambda)}{R_V}  & \quad  a_i > a_\mathrm{BC}
\end{cases}
\end{equation}

\noindent where $a_\mathrm{BC}$ is the lifetime of stellar birth-clouds as described in Section \ref{subsubsect:pipes_nebular}, $A_V$ is the attenuation in the $V$ band ($\sim5500$ \AA) in magnitudes, $\epsilon$ is a constant which can be used to control the extra attenuation towards H\,\textsc{ii} regions and $k(\lambda)$ and $R_V$ are specific to the dust model being used.

Three dust models are currently implemented: the \cite{Calzetti2000} law for local star-forming galaxies, the \cite{Cardelli1989} Milky Way dust law, and a flexible model based on that of \cite{Charlot2000}. For the first two of these models the form of $k(\lambda)$ and the value of $R_V$ are specified in the paper cited. For use in \bagpipes, these are extrapolated to shorter wavelengths using a power-law fit to the dust curve in the near-ultraviolet. For the \cite{Charlot2000} model we define

\begin{equation}
\frac{k(\lambda)}{R_V} = (\lambda/5500\text{\AA})^{-n}
\end{equation}

\noindent where $n$ is the slope of the attenuation law. Our $\epsilon$ parameter is the reciprocal of the parameter $\mu$ often used in the literature (e.g. \citealt{Cullen2017a}) for models of this type.

Dust emission from the neutral ISM is currently modelled as a single-temperature greybody, with flux per unit frequency, $S_\mathrm{gb}(\nu)$ (e.g. \citealt{Hildebrand1983}; \citealt{Younger2009}) given  by

\begin{equation}
S_\mathrm{gb}(\nu) \propto \nu^{\beta} B_\nu (T) = \frac{2h}{c^2} \frac{\nu^{\beta + 3}}{e^\frac{hc}{\lambda kT} - 1}
\end{equation}

\noindent where $T$ is the temperature of the greybody, $\beta$ is the spectral emissivity index and the dust has been assumed to be optically thin. The normalisation of this component is set such that the total energy removed from the spectrum by dust attenuation is the same as the total energy emitted by the greybody.

Dust emission in \bagpipes\ is hence modelled by two separate components. Firstly, the hot-dust component included in the \textsc{Cloudy} output diffuse continua for H\,\textsc{ii} regions (described in Section \ref{subsubsect:pipes_nebular}), and secondly the greybody component due to cold, diffuse dust emission. This kind of two-component approach has been shown to be successful in modelling observed infrared SEDs (e.g. \citealt{Casey2012}), and work is ongoing to test our model against observational data at these wavelengths. In the future we hope to implement the option of a more complex physical model for dust emission.

\subsubsection{IGM attenuation}\label{subsubsect:pipes_igm}

For $T_{\mathrm{IGM}}(\lambda, z_\mathrm{obs})$, \bagpipes\ incorporates the IGM attenuation model of \cite{Inoue2014}, an updated version of the \cite{Madau1995} model. The analytic expression presented in their section 4 is calculated and tabulated for rest-frame wavelengths between 911.8 \AA\ and 1215.7 \AA, then interpolated by the code for use in model generation. We assume  $T_{\mathrm{IGM}}(\lambda, z_\mathrm{obs})$ = 0 for any flux at $\lambda < 911.8$ \AA, the majority of which is already removed when applying the nebular-emission prescription (see Section \ref{subsubsect:pipes_nebular}).

\subsubsection{Velocity dispersion and spectral sampling}\label{subsubsect:pipes_veldisp}

\bagpipes\ is designed to perform spectroscopic fitting as well as fitting to photometry, meaning the effects of velocity dispersion, $\sigma_\mathrm{vel}$, must be included in the model to match observed spectral features. To facilitate this, \bagpipes\ converts the wavelength sampling of all input SPS models to constant spectral resolution, $R = \frac{\lambda}{\Delta\lambda}$. By default, a low resolution of $R$ = 100 is used over regions which will be used only for calculation of output photometry, and a higher resolution of $R$ = 600 is used over regions which will be used to generate the output spectrum (in practice, $R$ is doubled to achieve Nyquist sampling at the resolutions quoted). The latter region is then convolved with a Gaussian kernel in velocity space to model the effect of velocity dispersion within the observed galaxy, which is assumed to be the same for the stellar and gas components.

\subsection{Model fitting}\label{subsect:pipes_fitting}

The second major aspect of \bagpipes\ is its ability to fit the models described in Section \ref{subsect:pipes_model} to observational data within the framework of Bayesian inference. These data can take the form of an observed spectrum and/or any number of photometry points, all of which can be fitted simultaneously by the code. In Section \ref{subsubsect:pipes_bayes} we outline the principles of our fitting method as applied to photometric observations, additional considerations when fitting spectroscopic observations will be addressed in a follow-up paper. In Section \ref{subsubsect:pipes_multinest} we outline how fitting is performed using the \textsc{MultiNest} nested sampling algorithm. Section \ref{subsubsect:pipes_priors} is a brief note on specifying prior distributions within \bagpipes.

\begin{figure*}
	\includegraphics[width=0.95\textwidth]{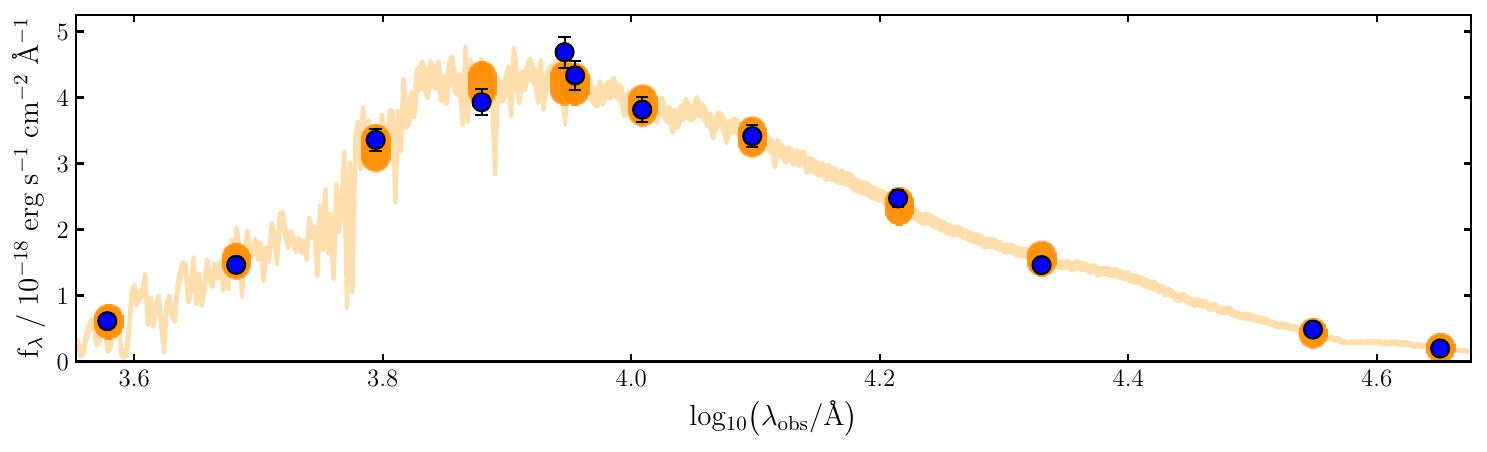}
	\includegraphics[width=0.99\textwidth]{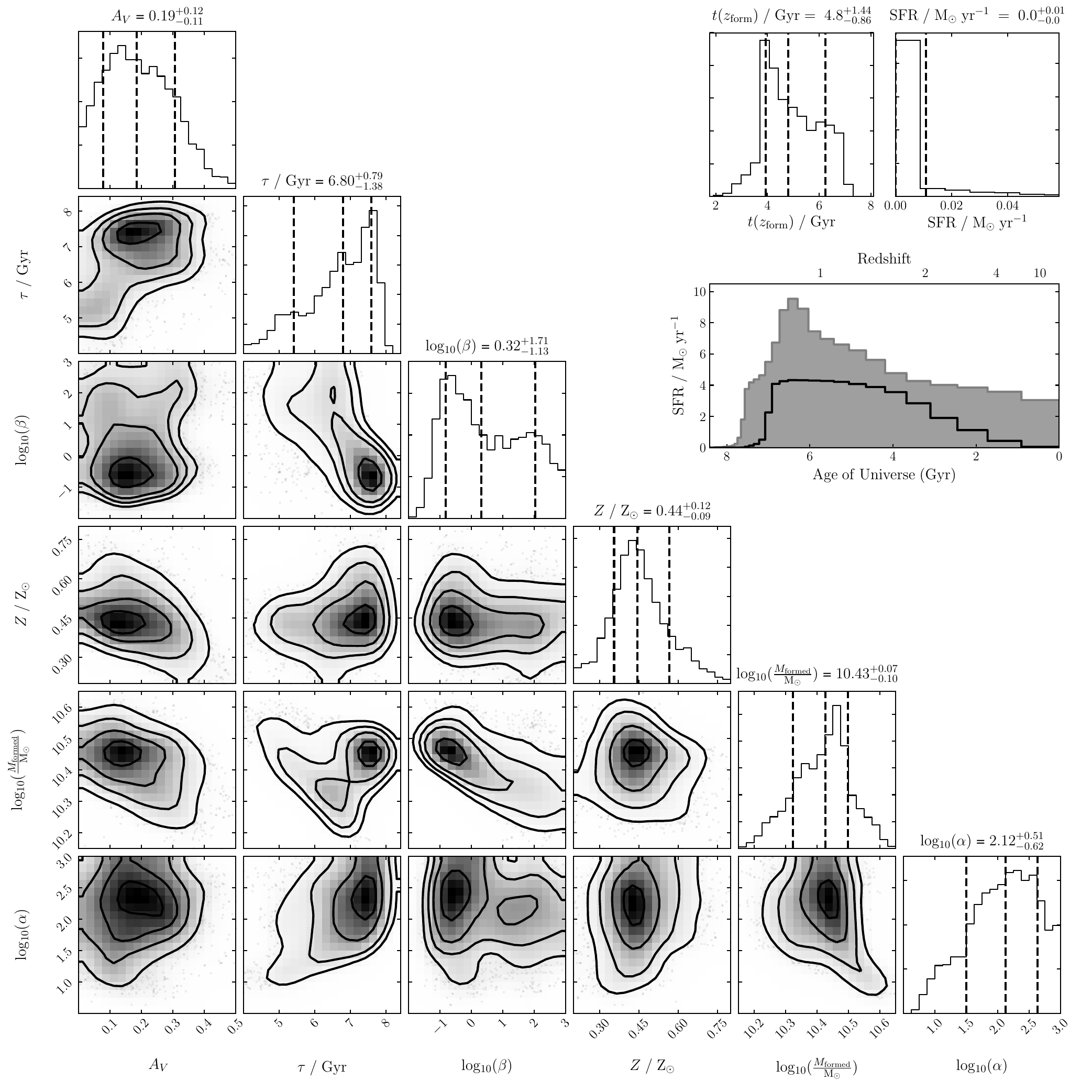}
    \caption{Example of \bagpipes\ output when fitting mock photometry from the \mufasa\ simulation for an object at at $z_\mathrm{obs} = 0.5$ using a double-power-law SFH. The mock photometry is shown on the top panel in blue (the bands are those of Table \ref{table:depths}). The 16th to 84th percentile range for the posterior spectrum and photometry are shaded orange. A corner plot showing the posterior for fitted parameters is shown below, with posterior SFH information shown to the right.}\label{fig:example_mufasa_obj}
\end{figure*}

\subsubsection{Bayesian inference methodology}\label{subsubsect:pipes_bayes}

Bayesian inference methods use Bayes' theorem to update prior knowledge about the probability of a hypothesis, or model, based on new data. This can take the form of parameter estimation, in which constraints on the parameters of a given model are updated, or model selection, in which the relative probabilities of two or more different models are assessed. Bayes' theorem states that for some new data, $\mathcal{D}$ and a hypothesis, $\mathcal{H}$ with parameter vector $\mathcal{\pmb{\Theta}}$,

\begin{equation}
P(\pmb{\Theta}\ |\ \mathcal{D},\ \mathcal{H}) = \frac{P(\mathcal{D}\ |\ \pmb{\Theta},\ \mathcal{H})\ P(\pmb{\Theta}\ |\ \mathcal{H})}{P(\mathcal{D}\ |\ \mathcal{H})}.
\end{equation}

\noindent Here, $P(\pmb{\Theta}\ |\ \mathcal{H})$ is the prior probability distribution for the model parameters: what is known before the new data are considered. $P(\mathcal{D}\ |\ \pmb{\Theta}, \mathcal{H}) = \mathcal{L}(\mathcal{D}\ |\ \pmb{\Theta},\ \mathcal{H})$ is the likelihood: the probability of obtaining the new data under the assumption of a certain set of parameter values. $P(\pmb{\Theta}\ |\ \mathcal{D},\ \mathcal{H}) $ is the posterior probability distribution: what is known once the new data have been considered. Finally, $P(\mathcal{D}\ |\ \mathcal{H})$ is the Bayesian evidence, or marginal likelihood: how good the hypothesis is overall at explaining the data.

For our purposes, the hypothesis consists of a certain model parameterisation within \bagpipes\ (e.g. type of dust model, number and type of SFH components, whether nebular emission is included etc). Once the model has been constructed, the user then defines prior probability distributions for its parameters (see Section \ref{subsubsect:pipes_priors} for details), and provides observational data, $f_i$ with associated uncertainties $\sigma_i$, which will be used to constrain those parameters.

The observational data are incorporated into the calculation through the likelihood function, which is constructed assuming that uncertainties are Gaussian and independent. When fitting only photometric data, the log-likelihood function used by \bagpipes\ is given by (e.g. \citealt{Hogg2010})

\begin{equation}\label{likelihood}
\mathrm{ln}\big(\mathcal{L}\big) = -0.5\sum_i \mathrm{ln}\Big( 2\pi \sigma_i^2 \Big)\ -0.5\sum_i \frac{\big(f_i - f_i^\mathcal{H}(\pmb{\Theta})\big)^2}{\sigma_i^2}
\end{equation}

\noindent where $f_i^\mathcal{H}(\pmb{\Theta})$ is the model prediction corresponding to the observed flux $f_i$, which is obtained by generating a \bagpipes\ model as described in Section \ref{subsect:pipes_model}.

\subsubsection{Nested Sampling implementation}\label{subsubsect:pipes_multinest}

\bagpipes\ makes use of the nested sampling algorithm of \cite{Skilling2006} to obtain the posterior distribution and evidence value given some model, prior distribution and observational data. Nested sampling is implemented in \bagpipes\ using \textsc{MultiNest} (\citealt{Feroz2008}; \citealt{Feroz2009, Feroz2013}), accessed through the \textsc{PyMultiNest} interface \citep{Buchner2014}.

Nested sampling allows for efficient exploration of higher-dimensional, multimodal and highly degenerate parameter spaces. This is invaluable in many circumstances relevant to spectral fitting, notably when dealing solely with broad-band photometric observations. In this case, the age-metallicity-dust degeneracy often leads to, at best, poorly constrained and highly degenerate parameter estimates, or at worst several widely spaced local minima in parameter space (e.g. \citealt{McLure2011}), any of which can trap traditional numerical functional minimisation routines or MCMC methods.

Once the posterior distribution is returned by \textsc{MultiNest}, it can be post-processed by \bagpipes\ to obtain other posterior information of interest, such as the posterior spectrum and SFH. A variety of visualisations, such as corner plots \citep{Foreman-Mackey2016} can also be constructed. An example corner plot and posterior spectral plot from a fit to mock photometry for one of the \mufasa\ objects introduced in Section \ref{sect:mufasa} is shown in Fig. \ref{fig:example_mufasa_obj}.

In the future, we intend to diversify the fitting options available in \bagpipes, giving the user access to more advanced versions of the nested sampling algorithm. This is necessary for faster and more reliable sampling of parameter spaces with higher dimensionalities, such as those encountered when fitting spectroscopic data. In particular we hope to include the option to implement dynamic nested sampling through the \textsc{Dynesty} package (Speagle et al. in prep).

\subsubsection{Specifying prior distributions on model parameters}\label{subsubsect:pipes_priors}

Any parameter which can be specified in the construction of a \bagpipes\ model, as described in Section \ref{subsect:pipes_model}, can be fitted using the code. When fitting a parameter, a prior probability distribution must be specified, consisting of an upper and lower limit on the parameter value, and a functional form for the prior probability density between these limits. In certain cases, these probability density functions will include hyperparameters which must also be specified. \bagpipes\ includes a separate ``priors'' module which allows the user to specify their own prior distributions, or pick from a number of options which have already been implemented.

\section{Testing Star-formation History Models with MUFASA}\label{sect:mufasa}

\bagpipes\ is a powerful tool for studying galaxy evolution, however, before applying it to real data it is important to test whether the models we define within it produce unbiased estimates of the parameters of interest. For example, it is often possible to obtain a good fit to photometric observations of a galaxy by modelling the whole SFH as a single burst of star formation (SSP model). This is likely to produce a reasonably accurate and unbiased estimate of, for example, the redshift of the galaxy. However, if this method were used to estimate the average time at which stars in the galaxy formed (see Section \ref{subsect:mufasa_sfh_comp}), it would clearly be biased towards younger ages, as the luminosities of galaxies are dominated by the youngest stars they contain. This is an example of model misspecification, where the chosen functional form for a model, or aspect of a model (in this case the SFH), is adequate for obtaining a fit to the data, however fails to accurately represent the underlying data-generating process, leading to biases in derived parameters.

Recently, \cite{Leja2017} performed tests to validate the \textsc{Prospector} inference framework by comparing physical properties derived from fitting only broad-band photometry against well-calibrated indicators derived from aperture-matched spectroscopy for a sample of local galaxies. They show that their model is capable of unbiased estimation of the SFR, dust-attenuation properties and stellar metallicities of their galaxies.

In this work, however, our focus is on the SFH properties of galaxies. This presents an additional challenge, as no independent, well-calibrated methods exist to obtain, for example, the average time at which the stars in a galaxy formed. Instead we look to simulations of galaxy formation to provide realistic SFHs which can be converted into mock photometric observations. These can then be fitted with \bagpipes\ using a variety of different model parameterisations with different priors, in order to find a model which accurately reproduces the properties of the input SFHs.

In this section we use the \mufasa\ suite of cosmological hydrodynamic simulations (see Section \ref{subsect:mufasa_mock}) to construct a catalogue of mock observations for simulated quenched galaxies with realistic SFHs, chemical-enrichment histories and dust properties. We tailor our mock observations to match as closely as possible the UltraVISTA data introduced in Section \ref{sect:uvista_data}. We then test the ability of a variety of models within \bagpipes\ to reproduce the SFH properties of quenched \mufasa\ galaxies.

We begin in Section \ref{subsect:mufasa_sfh_param} by discussing the advantages and disadvantages of different commonly used representations of the SFHs of galaxies. In Section \ref{subsect:mufasa_sfh_comp}, we introduce a set of parameters which can be used to compare how well different SFH models reproduce the important features of the input \mufasa\ SFHs. These parameters will also be used in the presentation of our results in Section \ref{sect:results}. We then give details of the \mufasa\ simulation and the process of generating mock observations in Section \ref{subsect:mufasa_mock}. Finally, in Section \ref{subsect:mufasa_fits} we fit these mock observations using \bagpipes, and compare the abilities of the exponentially declining and double-power-law SFH parameterisations to recover the SFH properties of \mufasa\ quenched galaxies. We demonstrate that, by using a double-power-law SFH model with the correct priors (see Table \ref{table:dblplaw_priors}), we can recover unbiased estimates of the underlying properties of the \mufasa\ SFHs. In contrast, we show that the exponentially declining model returns small but significant biases in these parameters. Based on these results, we proceed to use our double-power-law SFH model to fit the UltraVISTA dataset in Section \ref{sect:uvista_sample}.

\subsection{Common SFH parameterisations}\label{subsect:mufasa_sfh_param}

The traditional approach to SED fitting has been to use a simple functional form to parameterise the SFH, most commonly the exponentially declining function

\begin{equation}\label{tau}
\quad \mathrm{SFR}(t)\ \propto\ \begin{cases}
    \ \ \  e^{-\frac{t - T_0}{\tau}}             & \qquad  t > T_0\\
    \ \ \ 0 & \qquad  t < T_0
\end{cases}
\end{equation}

\noindent where $T_0$ and $\tau$ are parameters to be fitted. The main advantage of this rigid parameterisation is the speed of fitting, however \cite{Maraston2010} and \cite{Reddy2012} have shown by that it becomes less effective at higher redshifts, and \cite{Wuyts2011} and \cite{Pforr2012} have shown that the biases on the estimated SFH parameters (e.g. the ongoing SFR) are highly dependent on the permitted ranges of (alternatively, the priors on) model parameters.

It is possible to define more flexible parameterisations with the same number of parameters, such as the log-normal form of \cite{Gladders2013}, which has been shown by \cite{Diemer2017} to produce good fits to SFHs from the cosmological simulation Illustris \citep{Vogelsberger2014}. This form is clearly more physical, as star formation is required to increase smoothly from zero at the beginning of the Universe, rather than jumping from zero to its maximum value several billion years later. However, when using the log-normal form, star formation must always decline more slowly than it rises, meaning that this parameterisation struggles to model rapidly quenching systems.

Attempts have also been made to improve the flexibility of model SFHs by using more complex parameterisations, at the cost of increased computational expense. One option is the double-power-law form, which has previously been used to fit the evolution of the cosmic SFR density (e.g. \citealt{Behroozi2013}),

\begin{equation}\label{dblplaw}
\mathrm{SFR}(t)\ \propto\ \Bigg[\bigg(\frac{t}{\tau}\bigg)^{\alpha} + \bigg(\frac{t}{\tau}\bigg)^{-\beta}\Bigg]^{-1}
\end{equation}

\noindent where $\alpha$ and $\beta$ are the falling and rising slopes respectively, and $\tau$ is related to the time at which star formation peaks. The major advantage of the double-power-law SFH model is the decoupling of the rising and falling slopes of the SFH, allowing, in contrast to the log-normal form, a slow rise and rapid cutoff to star formation, as demonstrated in Fig. \ref{fig:example_pipes_model}. The double-power-law form is also discussed by \cite{Diemer2017} as a possible improvement on their log-normal parameterisation. They confirm that the double-power-law parameterisation produces better agreement with Illustris, particularly for rapidly quenching galaxies.

We also note that progress has been made in SED fitting using libraries of simulated SFHs from cosmological simulations (e.g. \citealt{Pacifici2012}), and using non-parametric SFHs (see Section \ref{subsubsect:pipes_sfh}). However these approaches are significantly more computationally intensive due to the inclusion of many free parameters, and are hence best suited to smaller sample sizes, rather than the large photometric catalogue we consider in this work. Testing of these methods within \bagpipes\ for smaller datasets is currently ongoing.

\subsection{Comparisons between SFH parameterisations}\label{subsect:mufasa_sfh_comp}

When attempting to compare results obtained using different SFH parameterisations, or to compare fitted SFHs with the input SFHs from \mufasa, it is necessary to define common parameters which can be derived from any SFH. One such parameter which is readily available is the living stellar mass present (excluding remnants) at the redshift of observation, $M_*(z_\mathrm{obs})$, or simply $M_*$.

To answer Question (i), posed in Section \ref{sect:introduction}, we require a method of quantifying when this mass was assembled. One common approach (e.g. \citealt{Thomas2017}) is to calculate the mass-weighted age, or in this case the mass-weighted time (measured forwards from the beginning of the Universe; see the note at the end of Section \ref{sect:introduction}) of a galaxy, $t_\mathrm{MW}$, given by

\begin{equation}\label{eqn:tform}
t_\mathrm{MW} = \frac{\int_{0}^{t_\mathrm{obs}} t\ \mathrm{SFR}(t)\ \mathrm{d}t}{\int_{0}^{t_\mathrm{obs}} \mathrm{SFR}(t)\ \mathrm{d}t} = t(z_\mathrm{form})
\end{equation}

\noindent where $t_\mathrm{obs} = t(z_\mathrm{obs})$. We then define the redshift of formation, $z_\mathrm{form}$, based on this mass-weighted time by setting $t_\mathrm{MW} = t(z_\mathrm{form})$.

In order to answer Question (ii), we require knowledge of the redshifts at which galaxies quenched, $z_\mathrm{quench}$. In order to define this we first require a definition of a quenched galaxy. This will also be necessary for selecting our quenched sample from the UltraVISTA catalogue. Two common methods exist for selecting samples of quenched galaxies, firstly based on their rest-frame UVJ colours \citep{Williams2009}, and secondly using a cut in specific SFR (sSFR; SFR divided by $M_*$), often evolving with observed redshift (e.g. \citealt{Pacifici2016}).

Relating the SFH of a galaxy to its rest-frame UVJ colours at earlier times requires several assumptions, such as the time-evolution of dust attenuation. This makes generalising UVJ selection to the past SFHs of galaxies extremely challenging. We therefore use a method similar to a specific SFR cut, by defining the dimensionless normalised SFR (nSFR), which is the current SFR as a fraction of the time-averaged SFR over the whole SFH, given by

\begin{equation}\label{passivity_param}
\mathrm{nSFR}(t) = \frac{\mathrm{SFR}(t)}{<\negmedspace\mathrm{SFR}\negmedspace>}\ =\ \mathrm{SFR}(t)\frac{t}{\int_{0}^{t} \mathrm{SFR}(t')\ \mathrm{d}t'}\ =\ \frac{t\ \mathrm{SFR}(t)}{M_\mathrm{formed}}
\end{equation}

\noindent where ${\int_{0}^{t} \mathrm{SFR}(t')\ \mathrm{d}t'} = M_\mathrm{formed}$ is the total mass of stars formed by the galaxy up to the time $t$, which is closely related to the living stellar mass.

\begin{figure}
	\includegraphics[width=\columnwidth]{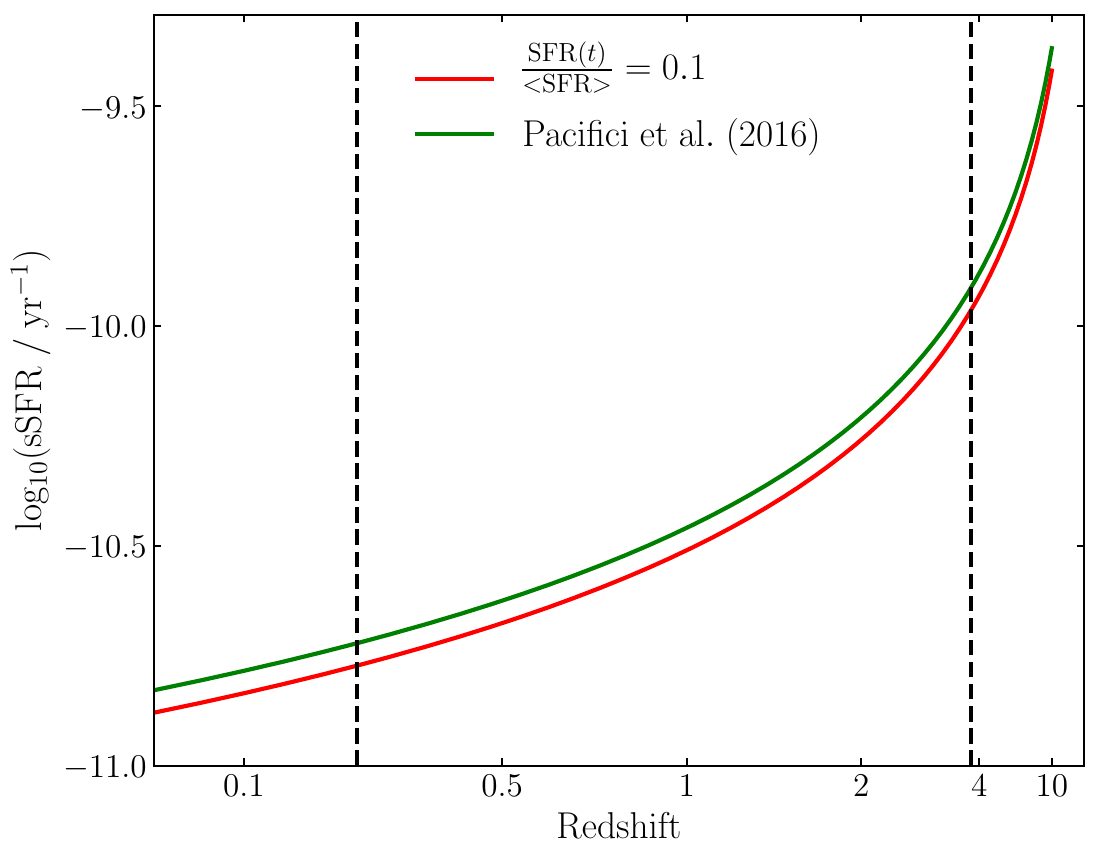}
	\caption{Selection criteria for quenched galaxies plotted as cuts in specific SFR as a function of redshift. Our cut in normalised SFR at 0.1 (see Section \protect \ref{subsect:mufasa_sfh_comp}) is shown in red, assuming that the living stellar mass, $M_*$ is 0.25 dex less than the total stellar mass formed (in reality this is dependent on the shape of the SFH). The selection criterion employed by \protect \cite{Pacifici2016} is shown in green. These can be seen to be very similar, and both produce good agreement with the UVJ selection criteria of \protect \cite{Williams2009}. The black dashed lines denote the edges of the redshift range spanned by our UltraVISTA sample.} \label{fig:ssfr_rsf}
\end{figure}

We define quenched galaxies as those which have normalised star-formation rate $< 0.1$, meaning the SFR at the time of observation must be less than 10 per cent of the average SFR across the history of the galaxy. This is effectively similar to an evolving cut in specific SFR, however it is not biased by the epoch at which the galaxy's stellar mass was assembled, as it depends on the total stellar mass formed rather than the living stellar mass. The value of 0.1 was chosen to preserve good agreement with UVJ selection (as demonstrated in Section \ref{sect:uvista_sample}). We show in Fig. \ref{fig:ssfr_rsf} how our cut in normalised SFR corresponds to a cut in sSFR and confirm that our selection criterion is very similar to that of \cite{Pacifici2016}, which they also find to be in good agreement with UVJ selection. We hence define $z_\mathrm{quench}$ as the redshift at which the normalised star-formation rate falls below 0.1. In the case of the \mufasa\ SFHs, which can be bursty and therefore fall below this threshold several times, we select the latest time at which this happens.

Finally, for our analysis in Section \ref{sect:results} it will be useful to define the quenching timescale, $\Delta t_\mathrm{quench}$ for our galaxies. We define this as being the duration between the time of formation $t(z_\mathrm{form})$, or mass-weighed time, and the time of quenching, $t(z_\mathrm{quench})$, such that

\begin{equation}
\Delta t_\mathrm{quench} = t(z_\mathrm{quench}) - t(z_\mathrm{form}).
\end{equation}

\noindent This definition of $\Delta t_\mathrm{quench}$ is designed to trace the timescale over which star-formation declines from its peak to a normalised SFR of 0.1, at which point we define the galaxy as quenched.

\begin{figure}
	\includegraphics[width=\columnwidth]{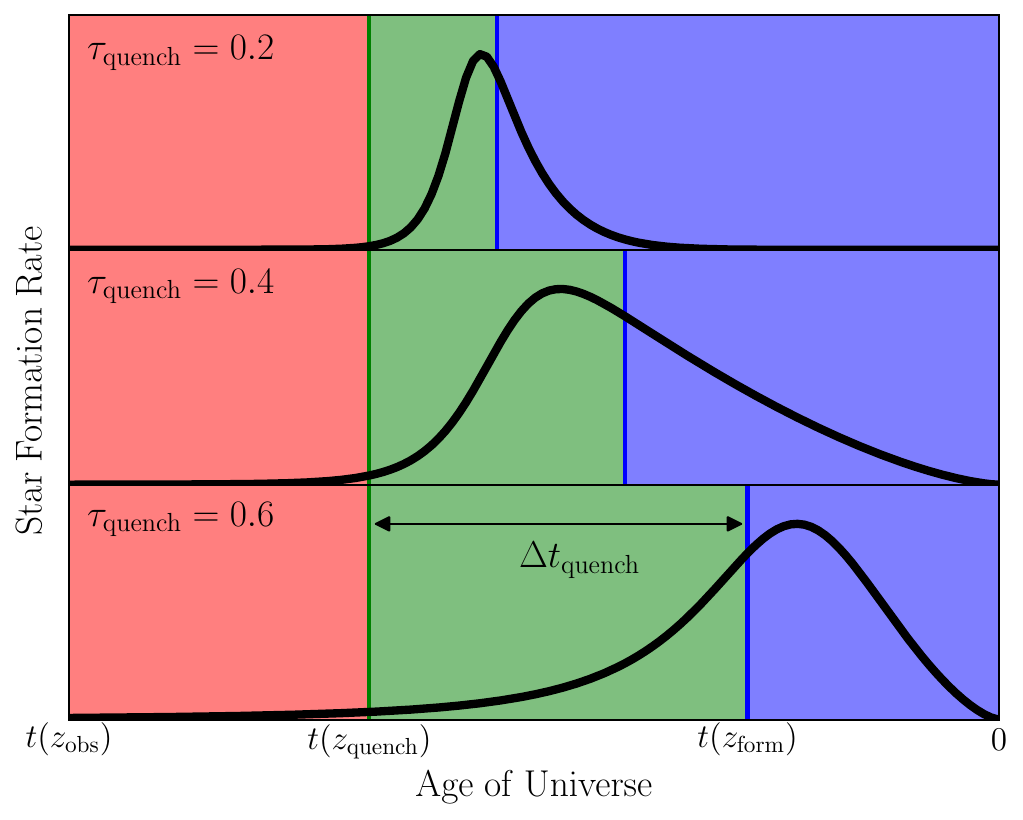}

	\caption{Pictorial representation of the scheme for describing and comparing galaxy SFHs introduced in Section \protect \ref{subsect:mufasa_sfh_comp}. The three example double-power-law SFHs shown all quench at the same time but have different shapes, leading to different values of $\tau_\mathrm{quench}$, which can be seen to trace both the speed of quenching, and, to a lesser extent, how quickly the SFH rises. The quenching timescale can be thought of as the time between peak star formation and the time of quenching.} \label{fig:example_mufasa_sfh}

\end{figure}

Figure \ref{fig:example_mufasa_sfh} shows how successful $\Delta t_\mathrm{quench}$ is in capturing this for a range of different SFH shapes. As can be seen, a rapidly quenched galaxy which has an extended rising wing will have a higher $\Delta t_\mathrm{quench}$ than one which quenches equally quickly, but also forms very rapidly. Thus $\Delta t_\mathrm{quench}$ traces both the speed of quenching as intended, but also has a secondary dependence on how extended the SFH is prior to the onset of quenching.

As the dynamical timescales of galaxies evolve with redshift, a parameter of significant interest is $\Delta t_\mathrm{quench}$ as a fraction of the age of the Universe when quenching occurs, $t(z_\mathrm{quench})$. This allows us to directly compare quenching timescales for galaxies at different observed redshifts. We define this normalised quenching timescale as $\tau_\mathrm{quench}$, where
\begin{equation}\label{eqn:f_shutoff}
\tau_\mathrm{quench} = \frac{\Delta t_\mathrm{shutoff}}{t(z_\mathrm{quench})}.
\end{equation}

\subsection{Generating mock observations of \mufasa\ galaxies}\label{subsect:mufasa_mock}

\mufasa\ \citep{Dave2016} is a suite of cosmological hydrodynamic simulations which correctly reproduces many of the observed properties of the population of quenched galaxies at $z=0$, including the slope of the red sequence and downsizing trend, as well as correct number densities for high-mass ($M_* \gtrsim 10^{10.7}\ \mathrm{M_\odot}$) quenched galaxies out to $z\sim2$ \citep{Dave2017}. This makes \mufasa\ an ideal resource for testing the ability of our model to recover realistic SFHs for massive quenched galaxies across a wide range of redshifts.

We use the post-processed data-products from \mufasa\ described by \cite{Dave2017}, which consist of a list of star particles for each galaxy with individual masses, ages, metallicities and $A_V$ values. We use snapshots of the simulation taken between $z = 0.5$ and 2.5 at intervals of $\Delta z = 0.5$.  To begin with, we select all galaxies from each of these snapshots with $M_*> 10^{10}\ \mathrm{M_\odot}$ in order to match the sample of observed galaxies we intend to extract from UltraVISTA (see Section \ref{sect:uvista_sample}).

For each \mufasa\ galaxy we generate a spectrum using \bagpipes\ by modelling each star particle as a burst of star formation. We use the BC03 models and the \cite{Calzetti2000} attenuation curve, which has been shown recently to be a robust average of the attenuation curves observed for galaxies across a wide range of redshifts (e.g. \citealt{Cullen2017b, Cullen2017a}; \citealt{McLure2017}; but see \citealt{Kriek2013}). We also fix log$_{10}(U)$ = $-3$ and $a_\mathrm{BC}$ to 10 Myr for our nebular-emission model, as these properties have only marginal impact on broad-band photometry for quiescent galaxies. For each galaxy we then calculate observed magnitudes in the twelve filters of the UltraVISTA catalogue described in Section \ref{sect:uvista_data}.

We also compile the simulated SFHs of these objects in 100 Myr bins and calculate the normalised SFR (Equation \ref{passivity_param}) at the time of observation using the SFR in the most recent 100 Myr bin as the current ongoing SFR. We then select objects with normalised SFR $< 0.1$ as our quenched sample, as described in Section \ref{subsect:mufasa_sfh_comp}. For the SFHs of our quenched sample we then calculate the three indicators described in Section \ref{subsect:mufasa_sfh_comp}, $M_*$, $z_\mathrm{quench}$ and $z_\mathrm{form}$.

Once the sample had been selected, Gaussian noise was added to the mock photometry at the levels listed for the UltraVISTA wide strips in Table \ref{table:depths}. An error floor was added such that no photometry point corresponded to a detection of greater than 20$\sigma$, or, in the case of the IRAC channels, 10$\sigma$. Finally a scatter was added to the catalogued observed redshifts of the \mufasa\ galaxies at the same level $(dz = 0.02)$ as the photometric redshifts calculated for our UltraVISTA catalogue. At the end of this process we had built a catalogue of mock observations for 677 simulated quenched galaxies, tailored to match as closely as possible the sample we intended to select from the UltraVISTA data of Section \ref{sect:uvista_data}.

\begin{figure*}
	\includegraphics[width=0.98\textwidth]{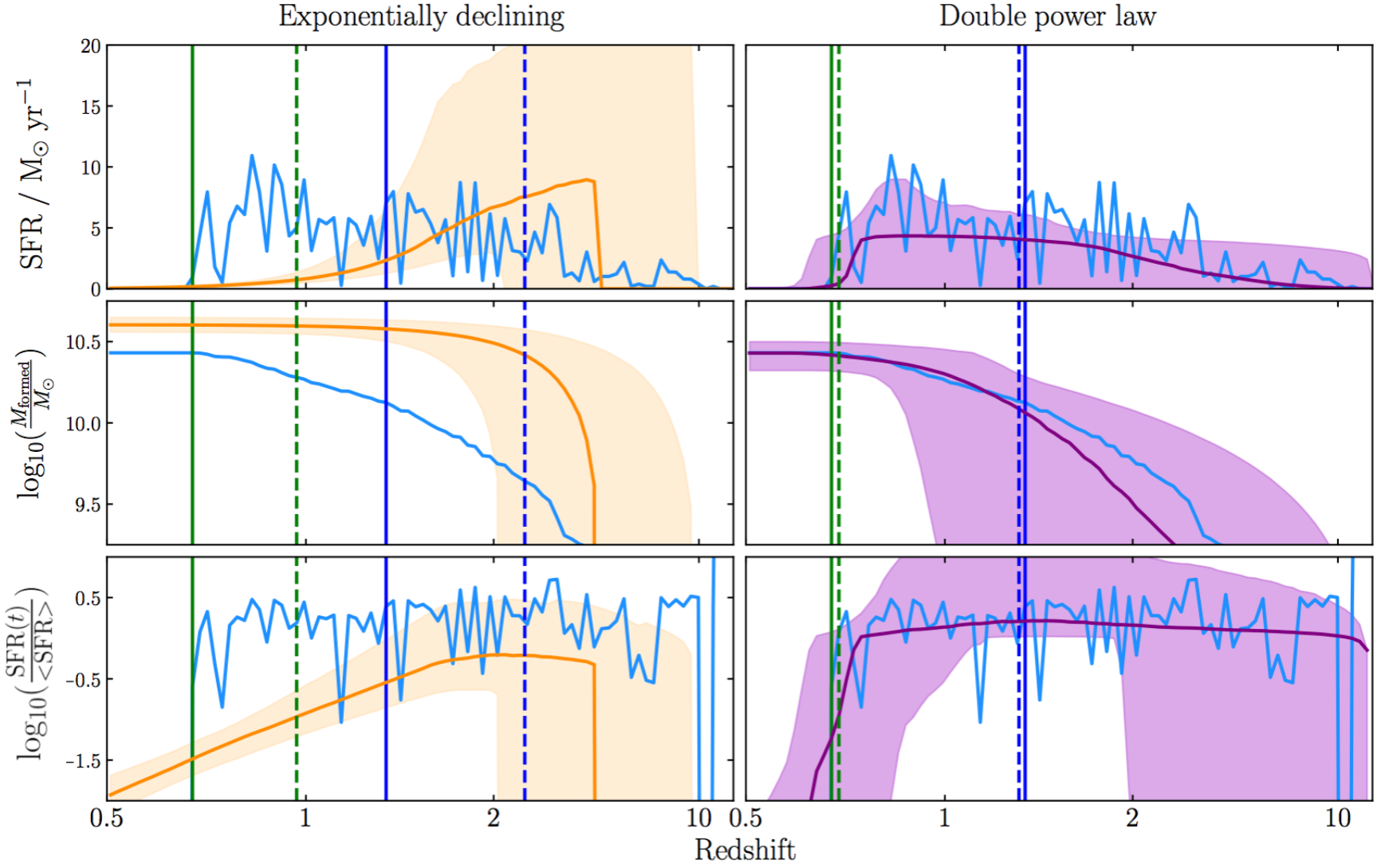}
	\caption{SFH (top), mass-assembly history (middle) and normalised SFR history (bottom) for the same example object from \mufasa\ as shown in Fig. \protect \ref{fig:example_mufasa_obj}. The input histories from \mufasa\ are shown in blue with the posteriors shown in orange (left) and purple (right) for the exponentially declining and double-power-law SFH models of Section \ref{subsect:mufasa_fits} respectively. The shaded regions show the 16$^\mathrm{th}$ -- 84$^\mathrm{th}$ percentiles of the posterior. The true redshifts of quenching and of formation (as defined in Section \ref{subsect:mufasa_sfh_comp}) are overplotted as green and blue solid vertical lines respectively, with corresponding posterior median fitted values shown as dashed lines. This object was chosen as an example as it is broadly representative of the median trends visible in Fig. \protect \ref{fig:mufasa_tau} and Fig. \protect \ref{fig:mufasa_dblplaw}.}\label{fig:mufasa_indiv_sfh}
\end{figure*}

\subsection{Recovering star-formation histories for \mufasa\ quenched galaxies}\label{subsect:mufasa_fits}

In this section we present the results of fitting the mock observations produced in Section \ref{subsect:mufasa_mock} with different SFH parameterisations. We are interested primarily in SFH recovery, and hence leave discussion of other physical parameters to future work. We evaluate our ability to accurately recover the \mufasa\ SFHs using the three indicators introduced in Section \ref{subsect:mufasa_sfh_comp}, $M_*$, $z_\mathrm{quench}$ and $z_\mathrm{form}$.

For all of our fitted models we assume Calzetti dust with fixed $\epsilon$ = 3, and fit $A_V$ with a uniform prior between 0 and 4 magnitudes. For our nebular model we fix $a_\mathrm{BC}$ = 10 Myr and $\mathrm{log_{10}}(U) = -3$. In principle all of these fixed parameters could be fitted with \bagpipes, however their marginal impact on observed photometry, particularly for quenched galaxies, and the computational expense of fitting large numbers of parameters led us to fix them to representative values, consistent with results from the literature (see Section \ref{subsect:pipes_model}).

We first build a baseline model of the kind commonly used in the literature (e.g. \citealt{Mortlock2017}). We parameterise the SFHs with a single exponentially declining component (Equation \ref{tau}) and use uniform priors on all of our SFH parameters. We set the limits of our priors to zero and 50 Myr less than $t(z_\mathrm{obs})$ for $T_0$, between 300 Myr and 10 Gyr for $\tau$ (following \citealt{Wuyts2011} in both cases), between 1 and $10^{13}\ \mathrm{M_\odot}$ in stellar mass formed, and between 0.2 and 2.5 $\mathrm{Z_\odot}$ in metallicity (Solar metallicity is assumed to take the value $\mathrm{Z_\odot} = 0.02$).

\begin{table*}
  \caption{Fixed and fitted parameters with their associated priors for the double-power-law SFH model used to fit the UltraVISTA catalogue. Priors listed as logarithmic are uniform in log-space. Observed redshift values were taken from the \protect \cite{Mortlock2017} catalogue (see Section \protect \ref{sect:uvista_sample}). Priors placed on parameters which are well constrained by the data (e.g. stellar mass) have little to no effect on the results obtained.}
\begin{tabular}{cccccccc}
\hline
Free Parameter & Prior & limits & & & & Fixed Parameter & Value \\

\hline
$A_V$  & Uniform & (0, 4) & & & & $a_\mathrm{BC}\ /\ \mathrm{Myr}$ & 10 \\

$\mathrm{log_{10}}(M_\mathrm{formed}\ /\ \mathrm{M_\odot})$  & Uniform & (1, 13) & & & & log$_{10}(U)$ & $-3$\\

$Z\ /\ Z_\odot$ & Uniform & (0.2, 2.5) & & & &  $\epsilon$ & 3 \\

$\tau \ /\ \mathrm{Gyr}$ & Uniform & (0, $t(z_\mathrm{obs}))$ & & & & $z_\mathrm{obs}$ & $z_\mathrm{M17}$ \\

$\alpha$ & Logarithmic & ($10^{-2}, 10^{3}$) & & & & SPS models & BC03 \\

$\beta$ & Logarithmic & ($10^{-2}, 10^{3}$) & & & & IMF & \cite{Kroupa2001} \\

\hline
\end{tabular}
\label{table:dblplaw_priors}
\end{table*}

Once posterior distributions had been obtained for our model parameters, we generated posterior SFHs and used these to calculate the posterior distributions of $M_*$, $z_\mathrm{quench}$ and $z_\mathrm{form}$ for each object. A representative comparison between \mufasa\ input and \bagpipes\ output from fitting our exponentially declining model to mock photometry is shown in the left-hand panels of Fig. \ref{fig:mufasa_indiv_sfh}. At the top the SFH is shown, in the middle the mass-assembly history, and at the bottom the history of the normalised SFR (see Equation \ref{passivity_param}). It can be seen that, for this object, this model over-predicts $M_\mathrm{formed}$, and hence $M_*$, and over-predicts both $z_\mathrm{quench}$ and $z_\mathrm{form}$. There are several obvious discrepancies in the histories when viewed as a whole, with the overall shape of the SFH being poorly reproduced by the model, the mass assembly occurring much earlier than in \mufasa.

\begin{figure*}
	\includegraphics[width=\textwidth]{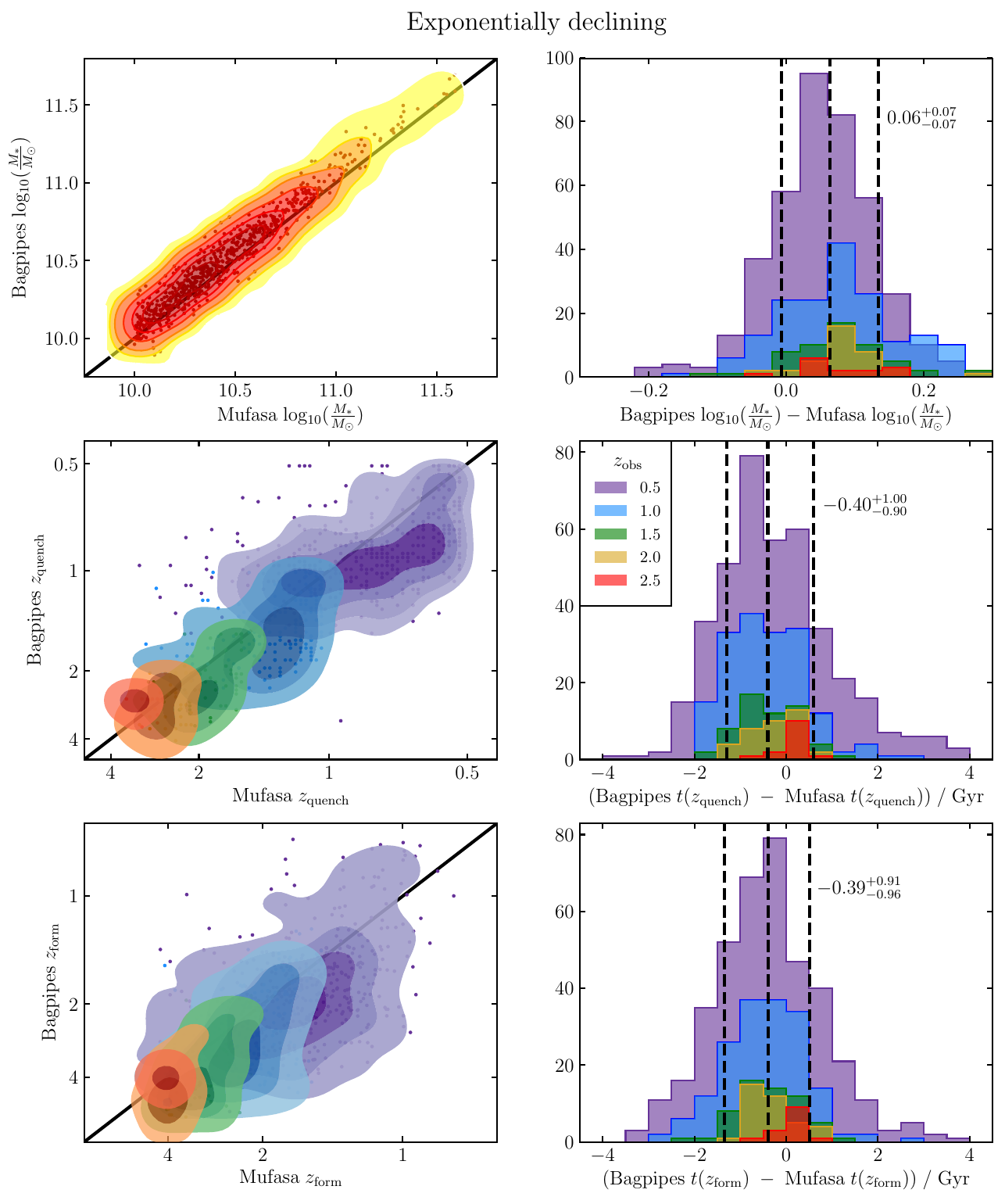}
	\caption{Comparison between SFH properties of \mufasa\ galaxies and posterior median values derived by fitting their mock photometric observations with \bagpipes\ using the exponentially declining SFH parameterisation described in Section \ref{subsect:mufasa_fits}. The top-left panel shows stellar mass, middle-left panel shows redshifts of quenching and bottom-left panel shows redshifts of formation for the whole \mufasa\ quenched sample of 677 objects. The right-hand panels show histograms of the offset between fitted and true values of log-stellar mass (top), time of quenching (middle) and time of formation (bottom). All panels except the top-left are split by colour for different observed redshifts, as shown on the middle-right panel. The dashed lines on the right-hand panels indicate the 16th, 50th and 84th percentiles of the combined distribution for all observed redshifts. Numerical ranges for these confidence intervals are shown on the figure. The stellar masses are overestimated by $\sim15$ per cent on average, and both $t(z_\mathrm{quench})$ and $t(z_\mathrm{form})$ are underestimated on average by $\simeq0.4$ Gyr.}\label{fig:mufasa_tau}
\end{figure*}

Fig. \ref{fig:mufasa_tau} shows a comparison between input values of of $M_*$, $z_\mathrm{quench}$ and $z_\mathrm{form}$ from \mufasa\ and the posterior median values obtained by fitting our exponentially declining model to the whole of the \mufasa\ quenched sample. It can be seen that biases are present in the estimates of our SFH parameters. This is most notable for the estimated stellar masses, which are over-predicted on average by 0.06 dex ($\sim15$ per cent), with the posterior median stellar masses of $\sim80$ per cent of objects being overestimates. Whilst this is a relatively small bias compared to other uncertainties which exist in the determination of stellar masses (e.g. \citealt{Mobasher2015}), it is nevertheless significant, and interesting given the prevalence of this method for estimating stellar masses in the literature.

Both $t(z_\mathrm{quench})$ and $t(z_\mathrm{form})$ are underestimated on average by $\simeq0.4$ Gyr. This is perhaps surprisingly good agreement, given that exponentially declining SFHs were not designed to reproduce these properties. However, as can be seen in the middle-left panel of Fig. \ref{fig:mufasa_tau}, there is considerable systematic variation in the degree of underestimation of $t(z_\mathrm{quench})$ with the time between $t(z_\mathrm{obs})$ and $t(z_\mathrm{quench})$, most notably at $z_\mathrm{obs} = 0.5$, with the quenching times of the earliest-quenching objects actually being, on average, overestimated. This is a problem as it results in inconsistencies between quenching properties derived for different sub-samples of objects.

Motivated by these observations, we tested other SFH parameterisations and combinations of priors in order to attempt to obtain better agreement with the true values of $M_*$, $z_\mathrm{quench}$ and $z_\mathrm{form}$ from \mufasa. We first investigated the effects of fitting the exponentially declining SFH parameterisation with different combinations of priors, but found no other combination that allowed us to obtain significantly better agreement with \mufasa.

We then considered the double-power-law form of Equation \ref{dblplaw}. To incorporate our prior expectation, based on the hierarchical growth of galaxies observed in cosmological simulations, that most SFHs are extended almost all the way back to the beginning of cosmic time, we impose a logarithmic prior on the rising slope, $\beta$. In the absence of constraining data, this produces an extended, gradual rise in the SFH up to the point at which the imprint of more recent star formation becomes obvious in the shape of the observed SED. We also put a logarithmic prior on the falling slope, $\alpha$, however this can be easily overcome by the strong imprint which recent, rapidly quenched star formation leaves on the observed SED.

\begin{figure*}
	\includegraphics[width=\textwidth]{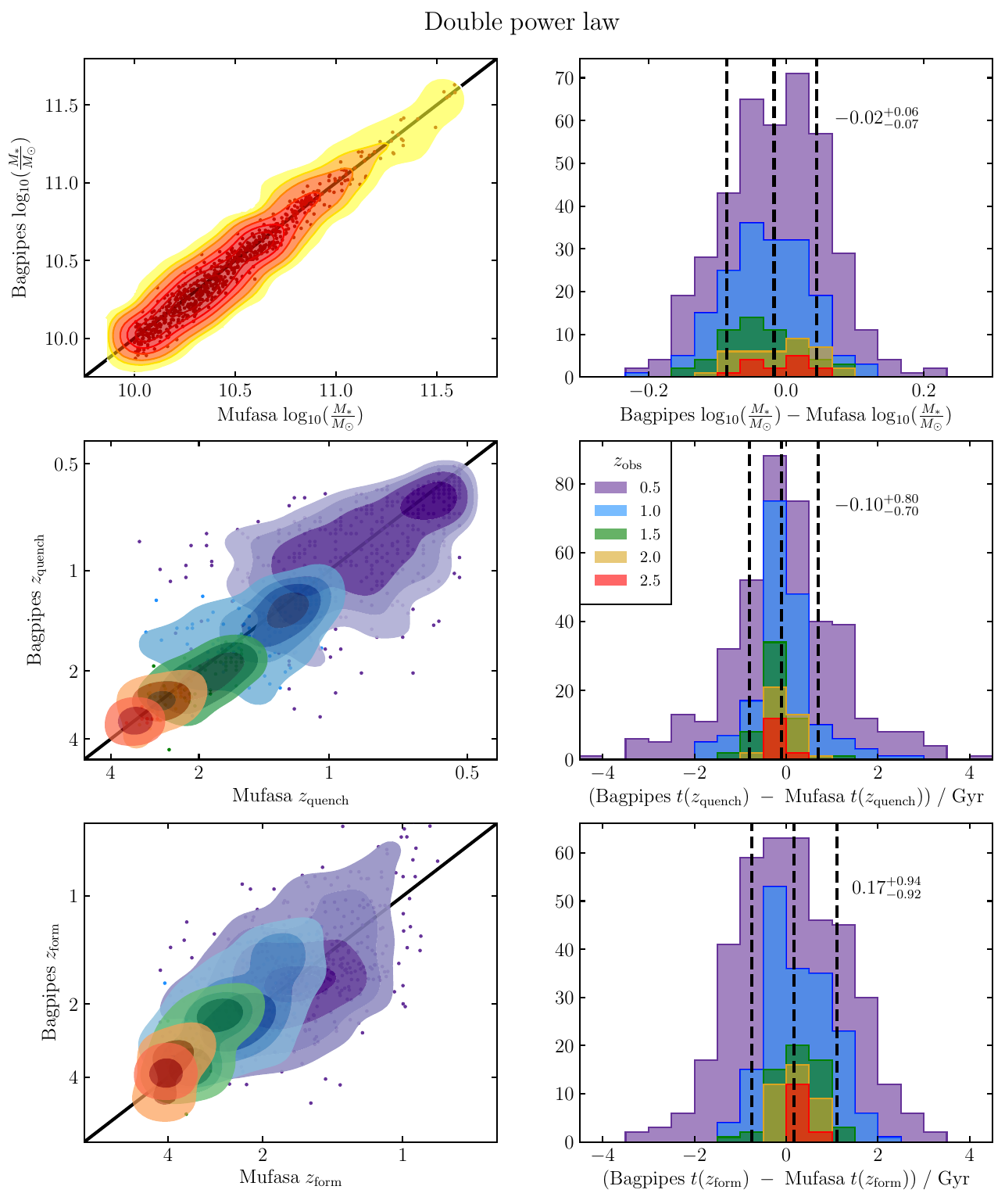}
	\caption{Comparison between SFH properties of \mufasa\ galaxies and posterior median values derived by fitting their mock photometric observations with \bagpipes\ using the double-power-law SFH parameterisation described in Table \protect \ref{table:dblplaw_priors}. Plot details are as in Fig. \protect \ref{fig:mufasa_tau}. Considerably better agreement can be seen for all three parameters at each observed redshift when compared to the exponentially declining SFH results displayed in Fig. \protect \ref{fig:mufasa_tau}.}\label{fig:mufasa_dblplaw}
\end{figure*}

\begin{figure*}
	\includegraphics[width=\textwidth]{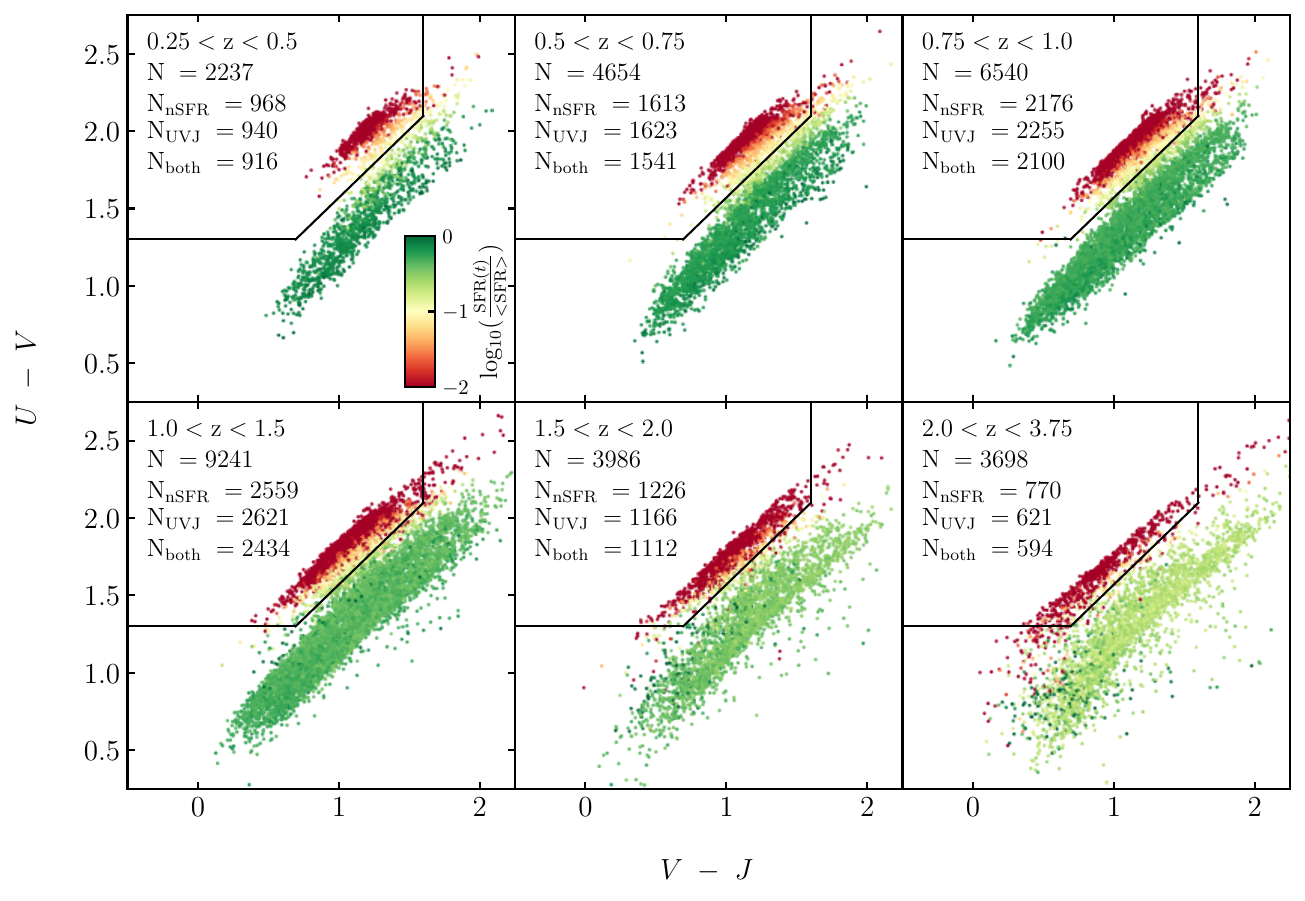}
	\caption{UVJ selection diagram for the sample of galaxies from UltraVISTA described in Section \ref{sect:uvista_sample}. The population of star-forming galaxies which match all of our criteria except our cut in normalised SFR (nSFR; see Equation \ref{passivity_param}) are also shown. A strong trend in nSFR perpendicular to the star-forming sequence can be seen, as has been observed for sSFR (e.g. \protect \citealt{Fang2017}). A UVJ selection of the kind proposed by \protect \cite{Williams2009}, with limits ($U - V$) > 1.3, ($V - J$) < 1.6, ($U - V$) > 0.88$\times$($V - J$) + 0.69 is overplotted in black for reference, however this was not included in our selection criteria. Our selection criterion for identifying quenched galaxies by $\mathrm{nSFR} < 0.1$ can be seen to be in good agreement with the UVJ selection shown in black at all observed redshifts.}\label{fig:uvista_UVJ}
\end{figure*}

\begin{figure}
	\includegraphics[width=\columnwidth]{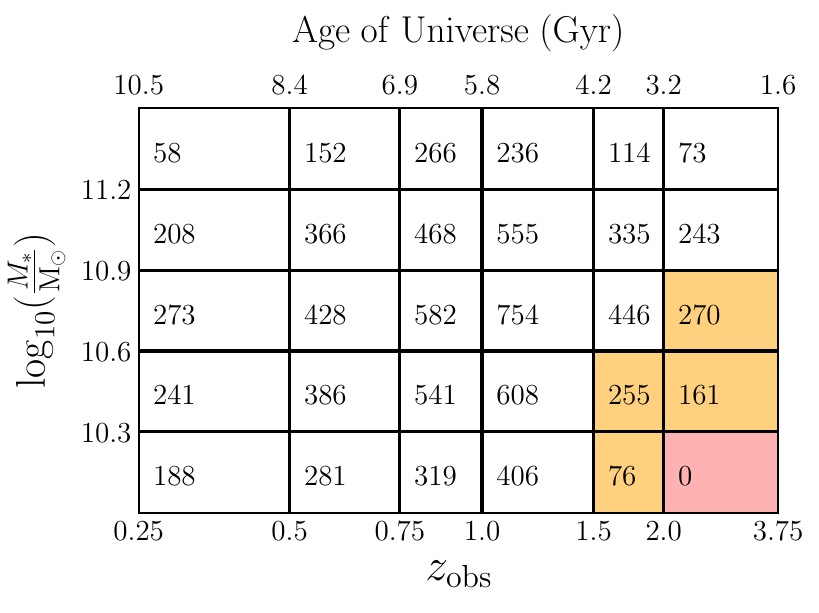}
	\caption{Number of galaxies in each of our bins in stellar mass and observed redshift. The edges of our bins are shown in black. Bins shaded orange are those for which UltraVISTA wide is not mass complete, and hence objects from the wide area were removed from our sample. The bin shaded red falls below the mass-completeness limit of both the wide and deep regions, and hence was excluded entirely from our sample.}\label{fig:zred_mass_dist}
\end{figure}

A list of the parameters and priors for our best double-power-law model are presented in Table \ref{table:dblplaw_priors}. Fig. \ref{fig:mufasa_dblplaw} shows the same comparison between input and fitted parameters for our double-power-law model as was shown for the exponentially declining model in Fig. \ref{fig:mufasa_tau}. It can be seen that considerably better agreement is achieved for all three of our SFH parameters, with stellar masses now offset by only 0.02 dex. The most significant improvement is in the estimation of $t(z_\mathrm{quench})$, where the bias on the median has been cut to 100 Myr. More significantly, the contours on the middle-left panel are now well aligned with the diagonal, indicating no change to the bias in our estimates of $z_\mathrm{quench}$ with increasing time interval between $t(z_\mathrm{obs})$ and $t(z_\mathrm{quench})$. Additionally, the observed scatter in our estimates of $t(z_\mathrm{quench})$ have been cut by $\simeq20$ per cent.

Fig. \ref{fig:example_mufasa_obj} shows example spectral and corner plots for a fit to one of the \mufasa\ objects using this model. This is the same object as shown in Fig. \ref{fig:mufasa_indiv_sfh}. Our fitted SFH, mass assembly history and normalised SFR history for our double-power-law model are shown on the right-hand panels of Fig. \ref{fig:mufasa_indiv_sfh}. These can be seen to be in considerably better agreement with the input histories from \mufasa\ than for the exponentially declining model, with close agreement being achieved with the input histories at all times.

The better agreement achieved with input \mufasa\ SFH parameters using our double-power-law model led us to take this model forward for use in fitting our UltraVISTA sample in Section \ref{sect:uvista_sample}. Under the assumption that the SFHs of real massive quenched galaxies look similar to those from \mufasa, we have shown that this model will allow us to obtain negligibly biased estimates of $M_*$, $z_\mathrm{quench}$ and $z_\mathrm{form}$ for the galaxies in our UltraVISTA sample.

\section{UltraVISTA Fitting and Sample Selection}\label{sect:uvista_sample}

In this section we describe the process of selecting our sample of 9289 massive quenched galaxies from the UltraVISTA catalogue introduced in Section \ref{sect:uvista_data}. We begin by fitting our double-power-law model within \bagpipes\ to every object in the catalogue, except for those which were identified as stars or potential AGN by \cite{Mortlock2017}. Parameters and priors for our model are listed in Table \ref{table:dblplaw_priors}. Fitting our \bagpipes\ model whilst allowing observed redshift to vary produces more scatter in the recovered photometric redshifts than \cite{Mortlock2017} obtained by taking the median result of five photometric-redshift codes, therefore it was decided to fix our observed redshifts to these median values.

We then select objects with posterior median stellar masses, $M_* > 10^{10}\ \mathrm{M_\odot}$ and calculate the posterior distributions of the normalised SFR (as defined in Section \ref{subsect:mufasa_sfh_comp}) for these objects. We then select our quenched sample to contain galaxies which have posterior median normalised SFR values less than 0.1. At this stage we also exclude poorly fit objects with minimum reduced chi-squared values of greater than 3. These objects are almost exclusively the result of individual datapoints affected by large systematic errors. They comprise $\sim1$ per cent of the sample and are not strongly clustered in observed redshift or stellar mass.

Fig. \ref{fig:uvista_UVJ} shows our final sample of quenched galaxies on the UVJ plane, as well as star-forming galaxies which match all of our other selection criteria except our cut in normalised SFR. A UVJ selection of the kind proposed by \cite{Williams2009} is also plotted. Our normalised SFR selection and UVJ selection can be seen to select very similar samples of objects at all redshifts, with typically $\gtrsim90$ per cent agreement between the two methods. Our method can be seen to be more subtle in removing star-forming galaxies close to the UVJ boundary.

We split our sample into six bins in observed redshift between $z=0.25-3.75$, chosen to span similar intervals in cosmic time. We then split each of these into five bins in stellar mass. Our first four mass bins have widths of 0.3 dex from $\mathrm{log_{10}}(M_*/\mathrm{M_\odot}) = 10-11.2$, whilst our highest mass bin includes all objects with $\mathrm{log_{10}}(M_*/\mathrm{M_\odot}) > 11.2$. Fig. \ref{fig:zred_mass_dist} shows the distribution of our final quenched sample across our bins in stellar mass and observed redshift. Bin edges are marked in black.

The \cite{Mortlock2017} catalogue is mass complete down to $\mathrm{log_{10}}(M_*/\mathrm{M_\odot}) = 10$ over the UltraVISTA wide region at $z \lesssim 1.5$, and over the deep region at $z \lesssim 2$.  We therefore exclude objects from our highest-redshift, lowest-mass bins if they fall below the mass-completeness limits of the UltraVISTA region they are drawn from. Bins are shaded orange in Fig. \ref{fig:zred_mass_dist} for objects in the wide region having been excluded, or red for all objects having been excluded. These final criteria leave us with a sample of 9289 objects.

\section{The Star-Formation Histories of UltraVISTA Quenched Galaxies}\label{sect:results}

In this section we present and discuss the results of our analysis of the SFHs of massive quenched galaxies from UltraVISTA. We analyse the SFHs using the scheme presented in Section \ref{subsect:mufasa_sfh_comp}, by calculating posterior SFHs for each of our objects and using these to calculate posterior distributions for $t(z_\mathrm{form})$, $t(z_\mathrm{quench})$, $\Delta t_\mathrm{quench}$ and $\tau_\mathrm{quench}$. The goal of our analysis is to address Questions (i) and (ii) posed in Section \ref{sect:introduction}. Hence, in Section \ref{subsect:results_when_formed} we consider Question (i): when did quenched galaxies form their stellar mass? Then, in Section \ref{subsect:results_how_long} we consider Question (ii): how long did the process of quenching take?

As can be seen from Fig. \ref{fig:example_mufasa_obj}, the SFHs of our individual objects are relatively poorly constrained. At this point, the question of how best to present and draw conclusions from the richness of the posterior information we have obtained naturally arises. The optimal approach to analysis of this dataset would be to construct a Bayesian hierarchical model in which the values of $z_\mathrm{form}$ and $z_\mathrm{quench}$ for each object were assumed to be drawn from some overall sample distribution with stellar-mass and observed-redshift dependencies, which could then be constrained by our whole dataset. However the computational expense of implementing this (e.g. using Gibbs sampling) was judged to be prohibitive. Instead we have elected to represent the posterior distribution for each object by the median values of relevant parameters, and to take advantage of our large sample size to make inferences about how these vary across the stellar-mass and observed-redshift bins plotted in Fig. \ref{fig:zred_mass_dist}.

\begin{figure*}
	\includegraphics[width=\textwidth]{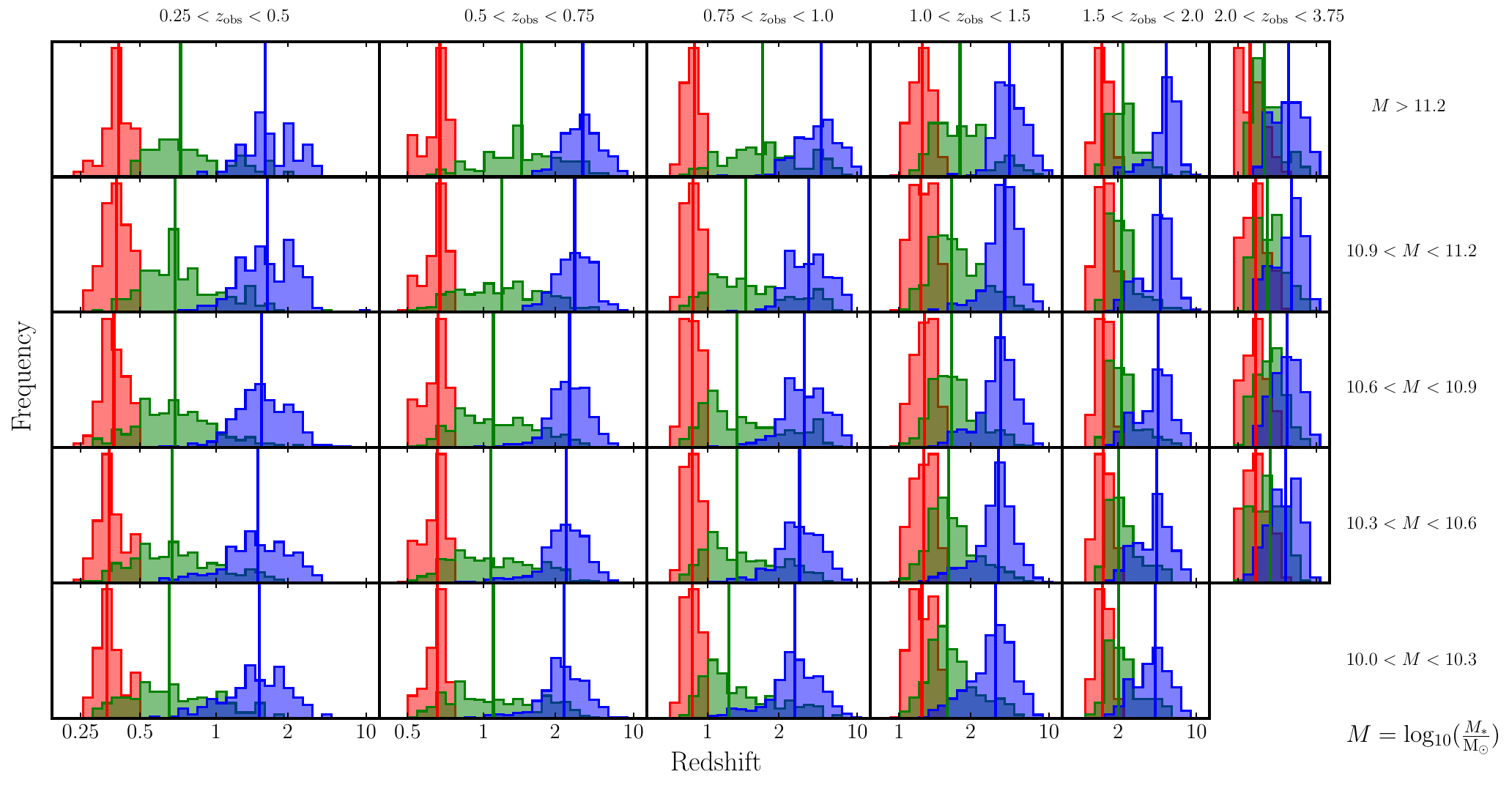}
	\caption{Histograms of the redshifts of observation (red), posterior median redshifts of quenching (green) and posterior median redshifts of formation (blue) for our sample of quenched galaxies, split into the stellar-mass and observed-redshift bins shown on Fig. \protect \ref{fig:zred_mass_dist}. The medians of the distributions for each bin are denoted by solid vertical lines. A clear downsizing trend can be seen in both the redshifts of formation and of quenching.}\label{fig:zobszqzform}
\end{figure*}

\subsection{When did quenched galaxies form their stellar mass?}\label{subsect:results_when_formed}

Fig. \ref{fig:zobszqzform} shows histograms of the posterior median values of $z_\mathrm{form}$ and $z_\mathrm{quench}$ for objects in our sample, as well as the distributions of observed redshifts, split into the stellar-mass and observed-redshift bins shown in Fig. \ref{fig:zred_mass_dist}. The distributions of $z_\mathrm{form}$ and $z_\mathrm{quench}$ can be seen to be, on the whole, unimodal and roughly Gaussian in shape. It is therefore reasonable to represent these distributions by their median values, which are plotted as solid vertical lines in Fig. \ref{fig:zobszqzform}. The median values of $z_\mathrm{form}$ with their standard errors are plotted against median stellar mass in Fig. \ref{fig:downsizing}.

Two clear trends are visible in the distributions of $z_\mathrm{form}$ from Fig \ref{fig:zobszqzform} and Fig. \ref{fig:downsizing}. Firstly there is the well-known downsizing trend, in which more massive galaxies at fixed observed redshift have higher median $z_\mathrm{form}$. This is discussed in Section \ref{subsubsect:downsizing}. Secondly, there is a trend towards progressively lower median $z_\mathrm{form}$ for objects at lower observed redshifts. We discuss this in Section \ref{subsubsect:results_zform_zobs_trend}.

\subsubsection{Downsizing: higher formation redshifts at higher masses}\label{subsubsect:downsizing}

The downsizing trend we observe, most clearly in Fig. \ref{fig:downsizing}, has also been observed in many other works (see Section \ref{sect:introduction}). \cite{Siudek2017} consider a sample of quenched galaxies at $0.4 < z < 1.0$ and show in their fig. 11 redshifts of formation calculated from stacked spectra in bins of $\Delta z = 0.1$ across the same range of stellar masses we consider. Their redshifts of formation are considerably lower than those we show in Fig. \ref{fig:downsizing} for our corresponding redshift bins. However, as they note, their methodology of assuming the SFHs of their galaxies to be composed of only a single burst component means their redshift of formation traces the time of the most recent episode of significant star formation. This means the formation redshifts they report are more analogous to our $z_\mathrm{quench}$ than our $z_\mathrm{form}$. It can be seen from Fig. \ref{fig:zobszqzform} that the median $z_\mathrm{quench}$ we derive for our $0.5 < z < 0.75$ and $0.75 < z < 1.0$ bins are consistent with the redshifts of formation shown on their fig. 11, ranging from $z \simeq 1.1$ at $M_* \sim 10^{10}\ \mathrm{M_\odot}$ to $z \simeq 1.5$ at the highest masses. It can also be seen from fig. 12 of \cite{Siudek2017} that these results are broadly consistent with analyses by \cite{Moresco2010}, \cite{Jorgensen2013} and \cite{Onodera2015}, whereas the results of \cite{Thomas2010} and \cite{Choi2014} confirm the continuation of the trends we observe in Fig. \ref{fig:downsizing} down to $z \simeq 0$.

\cite{Pacifici2016} perform an analysis more similar to ours than any of the studies mentioned above, both in terms of sample selection (see Fig. \ref{fig:ssfr_rsf}) and SFH recovery method, with the median SFHs they calculate also found to be well fitted by a double-power-law. Their redshifts of peak star formation, $z_\mathrm{peak}$ are therefore reasonably analogous to our $z_\mathrm{form}$. In their lowest-redshift bin $(0.2 < z < 0.5)$ they find a range of $z_\mathrm{peak}$ between $\simeq0.85$ at $M_* \sim10^{10}\ \mathrm{M_\odot}$ and $\simeq1.3$ at the highest masses; this is lower than the range of $z\simeq1.5$ to $1.7$ we find at similar observed redshifts in Fig. \ref{fig:downsizing}. This discrepancy is still apparent, though less marked at higher observed redshifts, up until $z_\mathrm{obs} \simeq 1.3$, after which our results are in good agreement. This disagreement is a consequence of differences in how the two redshifts are measured, with our $z_\mathrm{form}$ being higher than $z_\mathrm{peak}$ for SFHs which decline more rapidly than they rise (e.g. the middle panel of Fig. \ref{fig:example_mufasa_sfh}). The majority of our galaxies do follow this pattern (see Section \ref{subsect:results_how_long}) and so have $z_\mathrm{form}$ higher than $z_\mathrm{peak}$.

The downsizing trend we observe is relatively weak, with only $\simeq0.5$ Gyr in $t(z_\mathrm{form})$ separating the lowest and highest-mass galaxies in our sample at all observed redshifts. \cite{Pacifici2016} tentatively report a stronger downsizing trend at lower observed redshifts. This is not obvious in our results, with the strength of our downsizing trend across our whole observed redshift range being similar to that which they see at higher observed redshifts. However, in our lowest-redshift bin our number counts are significantly lower than at higher redshifts, increasing the statistical uncertainties on our results, as can be seen in Fig. \ref{fig:downsizing}.

In summary, our UltraVISTA SFHs exhibit a downsizing trend broadly consistent with results from the literature. Our trend is not particularly strong and does not evolve significantly with observed redshift. Our median $t(z_\mathrm{form})$ evolves by $\simeq0.5$ Gyr over a range of $\sim1.25$ dex in stellar mass, meaning the trend weakens as a fraction of the age of the Universe with decreasing observed redshift.

\subsubsection{Decreasing formation redshift with decreasing observed redshift}\label{subsubsect:results_zform_zobs_trend}

\begin{figure}
	\includegraphics[width=\columnwidth]{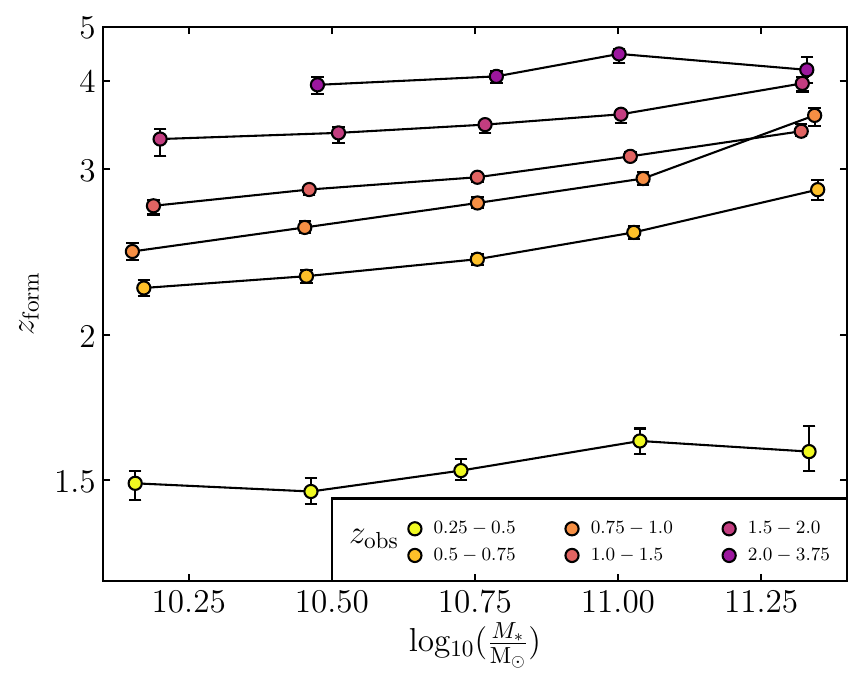}
	\caption{Median $z_\mathrm{form}$ for each of our mass and redshift bins with their standard errors. Two clear trends are visible. Firstly, at fixed observed redshift, galaxies with higher stellar masses are seen to have higher median $z_\mathrm{form}$. This is the commonly observed downsizing trend. Secondly, the median redshift of formation decreases with observed redshift at all masses. }\label{fig:downsizing}
\end{figure}

Along with the downsizing trend discussed in Section \ref{subsubsect:downsizing}, another common observation from similar analyses has been a trend towards lower average redshifts of formation with decreasing observed redshift for quiescent galaxies of all masses. This effect is observed for both $z_\mathrm{form}$ and $z_\mathrm{quench}$ in Fig. \ref{fig:zobszqzform}, and is clearly seen for $z_\mathrm{form}$ in Fig. \ref{fig:downsizing}. The effect is strongest at low redshift, with our lowest observed-redshift bin being strongly offset from the others, and persists across our whole observed-redshift range.

The simplest interpretation of this observation (as suggested by both \citealt{Pacifici2016} and \citealt{Siudek2017}) relates to the continuing assembly of the red sequence throughout our observed-redshift range (e.g. \citealt{Faber2007}; \citealt{Tomczak2014}). This means that new and younger galaxies are quenching throughout our observed-redshift range. Therefore our median $z_\mathrm{form}$ and $z_\mathrm{quench}$ values at low redshift are averages over a diverse population which quenches at a wide range of redshifts. Conversely, the quenched galaxies in our higher observed-redshift bins are a biased sub-sample of these, as our selection necessarily excludes high redshift star-forming galaxies  which will have quenched by lower redshifts.

The histograms plotted in Fig. \ref{fig:zobszqzform} support this picture, with the distributions of $z_\mathrm{form}$ and $z_\mathrm{quench}$ observed to be far more extended in our lower-observed-redshift bins. The fact that the distributions of $z_\mathrm{quench}$ are extended all the way down to the low-redshift ends of each of our observed-redshift bins is consistent with the assembly of the red sequence persisting throughout our observed-redshift range.

Another possible contributory factor to this trend of decreasing median $z_\mathrm{form}$ with decreasing observed redshift would be continuing evolution of the high-redshift quiescent population post-quenching, either through periods of rejuvenated star-formation, or merger events. This would mean that even if no galaxies joined the quiescent population across our observed redshift range, the median $z_\mathrm{form}$ we calculate at lower observed redshifts would still be lower than that which we observe at $z_\mathrm{obs} > 2$.Whilst our analysis confirms the well-known result that new galaxies join the red sequence across our observed redshift range (we see a strong increase in the number density of quiescent galaxies), we can also gain an insight into the roles of rejuvenated star-formation episodes and mergers.

A simple way of doing this is to compare the number density of quiescent galaxies we observe at high redshift (e.g. $z_\mathrm{obs} > 1.5$) with the number density of extremely old galaxies we observe at $0.25 < z_\mathrm{obs} < 0.5$ (those with $z_\mathrm{quench} > 1.5$). If all galaxies which are quenched at $z > 1.5$ undergo purely passive evolution from that time onwards, we would expect to observe the  same number density of objects in our $0.25 < z_\mathrm{obs} < 0.5$ bin with $z_\mathrm{quench} > 1.5$ as we find for quiescent galaxies at observed redshifts above 1.5.

However, if galaxies which are quenched at $z > 1.5$ undergo further evolution after this time, we should observe fewer extremely old objects at low redshift, as newer stellar populations have been added to objects which were already quenched at $z > 1.5$ since that time, decreasing our measured $z_\mathrm{quench}$ for these objects.

We restrict this comparison to objects with $\mathrm{log_{10}}(M_*/\mathrm{M_\odot}) > 10.3$, so that we have a mass-complete sample across our whole redshift range. Of the 780 galaxies in this mass range with $z_\mathrm{obs} < 0.5$, just 27 have $z_\mathrm{quench} > 1.5$, a comoving number density of $4.7\pm0.9 \times 10^{-5}$ Mpc$^{-3}$. By contrast we have a comoving number density of $11.9\pm0.3 \times 10^{-5}$ Mpc$^{-3}$for quiescent galaxies of these masses at $z_\mathrm{obs} > 1.5$. This suggests that the majority ($61\pm8$ per cent) of massive galaxies which quench at $z_\mathrm{obs}> 1.5$ undergo significant further evolution post-quenching by $z = 0.5$. This is in agreement with recent results from IllustrisTNG \citep{Nelson2017}, who find significant mass growth of galaxies post-quenching. It should be noted that these uncertainties take into account only Poisson noise on the number of objects observed, and not the uncertainties in $z_\mathrm{quench}$ for individual objects.

\subsection{How long did the process of quenching take?}\label{subsect:results_how_long}

A key piece of information necessary to constrain the physical mechanisms by which galaxies quench their star formation is the timescale over which quenching takes place. Accurate recovery of SFHs is the ideal method for studying quenching timescales, however, as discussed in Section \ref{subsect:mufasa_sfh_param}, recovering physically realistic SFHs is challenging. Even if this can be achieved there are questions as to which parameters of the SFH are most representative of the quenching timescale, usually taken to mean the timescale over which the galaxy crosses the green valley, i.e. between an initial position on the star-forming main sequence and some threshold defining passivity.

\begin{figure*}
    \includegraphics[width=\textwidth]{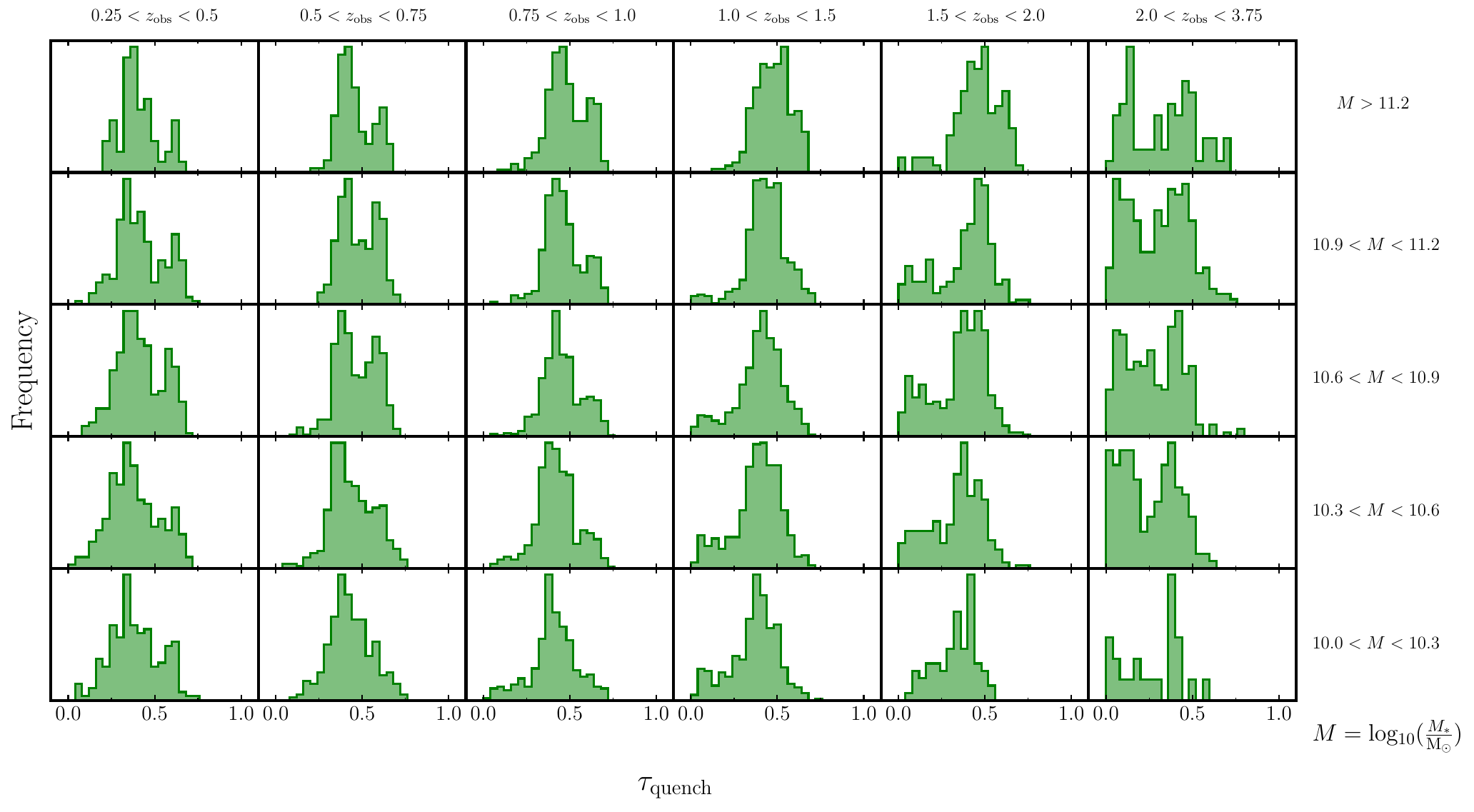}
    \flushleft
    \hspace*{0.4cm} \includegraphics[width=0.86\textwidth]{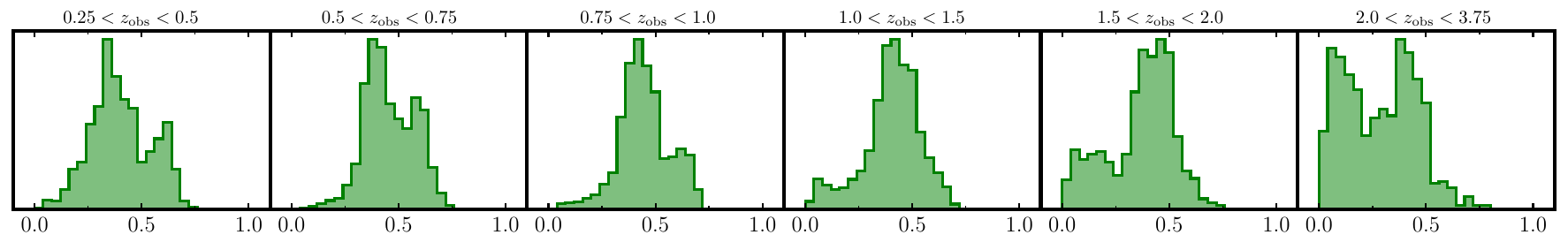}
    \caption{Distributions of posterior median values of $\tau_\mathrm{quench}$ (quenching timescale as a fraction of the age of the Universe when the galaxy quenched) across our bins in stellar mass and observed redshift. The bottom row shows the results from the grid collapsed along the stellar-mass axis for improved statistics.}\label{fig:tshutoff}
\end{figure*}

Recently, a common approach has been to fit SFH models similar to the exponentially declining model of Equation \ref{tau} to spectral indices such as D4000 and the strength of the H$\delta$ feature (e.g. \citealt{Nogueira2018}), or UV-optical colours derived from broad-band photometry (e.g. \citealt{Smethurst2017}). The timescale over which the SFR decreases, $\tau$ is then used as a proxy for quenching timescale. However, when fitting this kind of SFH model, strong degeneracies between $\tau$ and $T_0$ mean that constraining $\tau$ is extremely challenging. Also, the effects of the age-metallicity degeneracy on  D4000 and H$\delta$, and the age-metallicity-dust degeneracy on UV-optical colours must be properly treated, or the derived uncertainties in $\tau$ will be significantly underestimated.

In Section \ref{subsect:mufasa_sfh_comp} we define an alternative scheme, in which the quenching timescale, $\Delta t_\mathrm{quench}$ is the time interval between the formation redshift, $z_\mathrm{form}$ (corresponding to the mass-weighted age of the galaxy) and the redshift at which the SFR of the galaxy falls below 10 per cent of its average value across the SFH of the galaxy, $z_\mathrm{quench}$. We finally define the normalised quenching timescale, $\tau_\mathrm{quench}$ to be $\Delta t_\mathrm{quench}$ as a fraction of the age of the Universe at $z_\mathrm{quench}$. This is the fraction of the age of the Universe at the time of quenching which the galaxy takes to quench.

In Section \ref{subsect:mufasa_fits}, we confirm the ability of our double-power-law \bagpipes\ model to accurately recover both $z_\mathrm{form}$ and $z_\mathrm{quench}$. Hence we can be confident that, on average, we recover the correct values of $\tau_\mathrm{quench}$ for our UltraVISTA objects. This allows us to obtain information about the distribution of shapes their SFHs take and how these vary across the stellar-mass and observed-redshift bins shown in Fig. \ref{fig:zred_mass_dist}.

Fig. \ref{fig:tshutoff} shows the distributions of posterior median values of $\tau_\mathrm{quench}$ for objects in each of our stellar-mass and observed-redshift bins. Additionally, stacked histograms across all stellar-mass bins within a given observed-redshift bin are shown below the main grid. In Section \ref{subsubsect:results_how_long_fs_trends} we discuss trends in these histograms across our stellar-mass and observed-redshift bins, and what these reveal about trends in the shapes of the SFHs. In Section \ref{subsubsect:results_how_long_priors} we perform a check to understand the effects of our priors on these trends. In Section \ref{subsubsect:results_how_long_interp} we consider what these results reveal about the quenching mechanisms that have influenced the SFHs of the galaxies in our quenched sample.

\subsubsection{Trends in SFH shape revealed by $\tau_\mathrm{quench}$}\label{subsubsect:results_how_long_fs_trends}

Firstly, in our highest observed-redshift bin, $2.0 < z_\mathrm{obs} < 3.75$, there is a strong peak in the distribution of $\tau_\mathrm{quench}$ centred on $\tau_\mathrm{quench} \simeq 0.1$ in each of our mass bins. This peak is also apparent, though reduced in strength, across our two redshift bins between  $1.0 < z_\mathrm{obs} < 2.0$, and begins to take on a more obvious mass dependency, being observed most strongly at lower masses. A tail of small numbers of objects with $\tau_\mathrm{quench} < 0.2$ is apparent down to $z = 0.25$ at low masses. In contrast, virtually no objects are observed to fall into this region above $M_* = 10^{10.5}\ \mathrm{M_\odot}$ and below $z_\mathrm{obs} = 1$. A $\tau_\mathrm{quench}$ of 0.1 corresponds to a quenching timescale of $\sim300$ Myr at $z_\mathrm{quench} = 2$,  or $\sim600$ Myr at $z_\mathrm{quench} = 1$.

As demonstrated by the top panel of Figure \ref{fig:example_mufasa_sfh}, in order to reach such small values of $\tau_\mathrm{quench}$, SFHs must not only be rapidly quenched, they must also rise rapidly at late times, just before they quench. This shows that at $z_\mathrm{obs} \gtrsim 1$ there is a significant component of the quenched population with SFHs that both rise and fall rapidly, in a small fraction of the age of the Universe at their time of quenching. There is also a small component with this SFH shape at lower redshifts $(z \lesssim 1)$ in our lower-mass bins.

Secondly, across all of our bins in stellar mass and observed redshift there is a strong peak centred around $\tau_\mathrm{quench} \simeq 0.4$. This accounts for the bulk of objects in our sample, and indicates an extended period over which star formation rises, followed by a relatively abrupt shutoff (e.g. the middle panel of Fig. \ref{fig:example_mufasa_sfh}). These SFHs have quenching timescales of just under half the age of the Universe at their time of quenching (e.g. $\sim1$ Gyr at $z_\mathrm{quench} = 2$, or $\sim2$ Gyr at $z_\mathrm{quench} = 1$), almost as rapid as the previous category of objects. However they are distinct from the previous category as they undergo a much more extended period of rising star formation before they reach their peak.

Finally, a third peak begins to emerge at $z_\mathrm{obs} < 1.0$, centred around $\tau_\mathrm{quench} \simeq 0.6$. This population is more apparent at $0.5 < z_\mathrm{obs} < 1.0$ in our higher-mass bins, but by our lowest-redshift bin this peak is clearly visible even in our lowest-mass bin. These objects differ significantly from the previous two populations in that their SFHs rise more quickly than they fall (e.g. the bottom panel of Fig. \ref{fig:example_mufasa_sfh}), resembling the exponentially declining SFHs traditionally used in SED fitting, or the log-normal SFHs of \cite{Gladders2013}, but represent a minority of our observed sample. These objects have quenching timescales greater than half of the age of the Universe at $z_\mathrm{quench}$ (e.g. $\sim3$ Gyr at $z_\mathrm{quench} = 1$).

These trends are more clearly visible in the bottom row of Fig. \ref{fig:tshutoff}, where mass resolution has been sacrificed in order to obtain better statistics. The strongest peak at $\tau_\mathrm{quench} \sim 0.4$ is clearly dominant, whilst the presence and gradual decline of the population below $\tau_\mathrm{quench} = 0.2$ is clear at $z_\mathrm{obs} > 1$, and the presence and gradual rise of the peak centred on $\tau_\mathrm{quench} = 0.6$ is clear at $z_\mathrm{obs} < 1$.

Initially, Figure \ref{fig:tshutoff} appears to show surprisingly little variation in quenching timescale with stellar mass. However, it is important to remember that $\tau_\mathrm{quench}$ is normalised by the age of the Universe at $z_\mathrm{quench}$. As can be seen from Figure \ref{fig:zobszqzform}, more massive galaxies have higher median $z_\mathrm{quench}$ than less massive ones. This means that Figure \ref{fig:tshutoff} is actually showing more rapid quenching for more massive galaxies.

\subsubsection{A check on the effects of our priors}\label{subsubsect:results_how_long_priors}

Considerable effort has already been made in Section \ref{sect:mufasa} to understand the dependence of these results on the model we have used to fit our observed photometry. Here, we perform one further test to directly check for potential biasing effects of the priors of our SFH parameterisation on $\tau_\mathrm{quench}$. To do this we draw SFH parameters at random from the prior distributions listed in Table \ref{table:dblplaw_priors} and calculate $t(z_\mathrm{form})$, $t(z_\mathrm{quench})$ and $\tau_\mathrm{quench}$ for a statistical sample of SFHs.

\begin{figure*}
	\includegraphics[width=\textwidth]{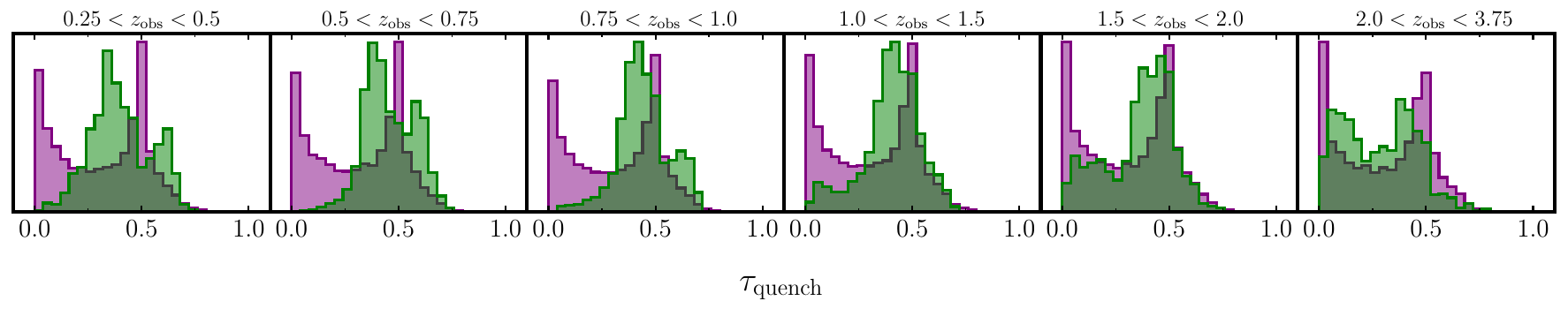}
	\includegraphics[width=\textwidth]{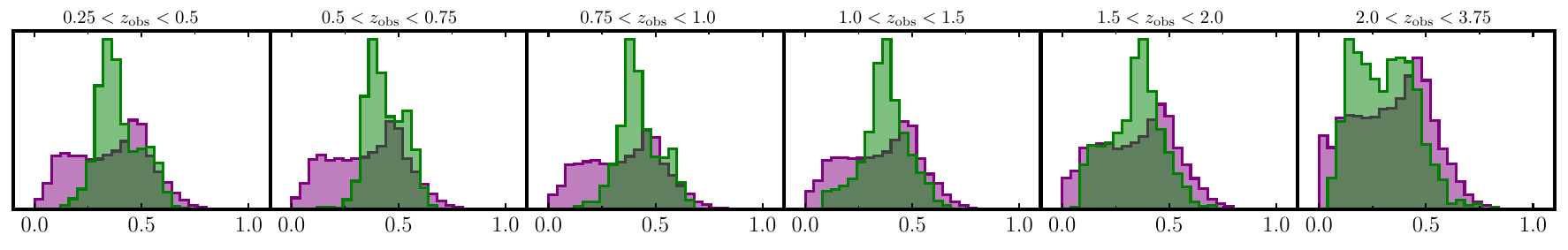}
	\caption{Distributions of $\tau_\mathrm{quench}$ in bins of observed redshift. The top row is the same as shown at the bottom of Fig. \protect \ref{fig:tshutoff} with the prior distribution shown in purple. The bottom row shows the prior distribution and posterior median $\tau_\mathrm{quench}$ values for the case in which the prior limits on $\alpha$ and $\beta$ have been reduced to (0.1, 100) from their original ranges of (0.01, 1000). This change, whilst flattening the prior distribution of $\tau_\mathrm{quench}$, does not significantly affect our results.}\label{fig:tshutoff_priors}
\end{figure*}

The value of $\tau_\mathrm{quench}$ depends only on the shape of the SFH, not its normalisation, therefore the prior distribution on $\tau_\mathrm{quench}$ is independent of stellar mass. On the top row of Fig. \ref{fig:tshutoff_priors}, our prior distributions are plotted in purple.  Only a very weak dependence on observed redshift is seen, meaning the trends we observe across our stellar-mass and observed-redshift bins cannot be a result of biasing effects of our prior distribution.

It is still important however to understand how the overall shape of our distribution is affected by our choice of priors. As can be seen from Fig. \ref{fig:tshutoff_priors}, our prior distribution has strong, narrow peaks at $\tau_\mathrm{quench}$ = 0 and 0.5. Upon experimentation it was found that these are due to our allowing $\alpha$ and $\beta$ for our double-power-law SFHs (see Equation \ref{dblplaw}) to reach very low and very high values respectively. Whilst these peaks in the prior do not correspond with any of the peaks discussed in Section \ref{subsubsect:results_how_long_fs_trends}, we still wished to understand how they might affect our results.

In order to check that these peaks do not strongly affect our results, we reduced the range of allowed $\alpha$ and $\beta$ values from (0.01, 1000) to (0.1, 100), which was found through experimentation to produce a relatively flat prior on $\tau_\mathrm{quench}$. We then re-fitted our sample using this reduced range of allowed $\alpha$ and $\beta$ values. This new prior distribution is shown in purple on the bottom row of Fig. \ref{fig:tshutoff_priors}, along with the results which were obtained by fitting with this new prior. It can be seen that the structure described in Section \ref{subsubsect:results_how_long_fs_trends} is still clearly visible under this new, flattened prior.

\subsubsection{Physical interpretation of trends in SFH shape}\label{subsubsect:results_how_long_interp}

To briefly recap Section \ref{subsubsect:results_how_long_fs_trends}, we observe three peaks in the distributions of $\tau_\mathrm{quench}$ in Fig. \ref{fig:tshutoff}, corresponding to three distinct SFH shapes. We refer to these three cases as (a), (b) and (c), which correspond roughly to the top, middle and bottom SFHs shown in Fig. \ref{fig:example_mufasa_sfh} respectively. To summarise their properties, we have:

\vspace{3mm}

\noindent \textbf{(a)} A peak centred on $\tau_\mathrm{quench} \simeq 0.1$, corresponding to SFHs which rise and decline rapidly, in a small fraction of the age of the Universe at $z_\mathrm{quench}$. Strongly visible at $z_\mathrm{obs} > 1$, also present in lower-mass bins at lower redshift. Physical quenching timescales of $< 1$ Gyr.

\vspace{3mm}

\noindent \textbf{(b)} A peak centred on $\tau_\mathrm{quench} \simeq 0.4$, corresponding to SFHs which decline more rapidly than they rise. Strongly visible in all of our bins of stellar mass and observed redshift, making up the majority of our observed sample. Physical quenching timescales of $1-2$ Gyr.

\vspace{3mm}

\noindent \textbf{(c)} A peak centred on $\tau_\mathrm{quench} \simeq 0.6$ corresponding to SFHs which rise more rapidly than they decline. Only visible at $z_\mathrm{obs} < 1$ and most visible at higher masses, making up a further minority of our observed sample. Physical quenching timescales of $\gtrsim3$ Gyr.

\vspace{3mm}


\noindent As these SFH types are seen to be clearly distinct from each other in Fig. \ref{fig:tshutoff}, it is reasonable to try to associate them with qualitatively different quenching processes. This is possible through comparing the physical quenching timescales we infer for each of these cases to predictions for the behaviour of different physical processes.

Case (a) is associated with a process which rapidly quenches star formation, on timescales of $< 1$ Gyr. This process is most active at $z_\mathrm{obs} > 1$, and is also associated with a rapid rise in star-formation activity immediately preceding quenching. However, these SFHs need not necessarily be the result of a monolithic-collapse scenario. As discussed in Section \ref{subsect:mufasa_fits}, more recent star formation leaves a more significant impact on the observed spectrum. Therefore SFHs with a slow, shallow rise from the beginning of cosmic time, followed by a large, rapidly rising and quenching burst will be fitted with something like case (a) if a double-power-law form is used. This is because the more recent star-formation episode dominates the observed spectrum. More strongly constraining data, such as rest-frame ultraviolet spectroscopy, would be needed to capture the detail of the early-time evolution of such SFHs.

The speed of quenching in case (a) is indicative of an ejective quenching process (see Section \ref{sect:introduction}), as shutting off the supply of new gas would not be sufficient to cause the almost immediate cessation of star-formation activity. Quasar-mode feedback is one process which has the capability to cause extremely rapid quenching, and, if triggered by major-merger events, would also be expected to be much more common at high redshift, as the rate of major mergers is known to be a strong function of redshift (e.g. \citealt{Lotz2011}). The merger-triggered, quasar-mode feedback scenario would also explain the rapid rise of star-formation activity which is observed immediately before these galaxies quench.

 Virtually no examples of this kind of SFH are observed in our lowest-redshift bins, except at the lowest masses, supporting the conclusion we reach in Section \ref{subsubsect:results_zform_zobs_trend} that the majority of galaxies which are observed to be quenched at high redshift undergo further interactions which affect the shapes of their SFHs when they are observed later in cosmic time. This is consistent with the findings of \cite{Gabor2011}, who conclude that systems which quench due to AGN-driven outflows should begin to re-accrete gas and resume star formation within $1{-}2$ Gyr.

The tidal/ram-pressure stripping experienced by galaxies falling into clusters is another process which can generate SFHs with this shape (e.g. \citealt{Diemer2017}). This could be responsible for the tail of objects at low masses at $z_\mathrm{obs} < 1$ with this kind of SFH, as the most massive galaxies are unlikely to become satellites, and mergers are less common later in cosmic time.

Case (b) is the dominant mode of quenching we observe within our sample, and is associated with a process which quenches star formation on intermediate timescales of $\sim 1{-}2$ Gyr. The dominant mode of quenching in modern hydrodynamic simulations such as \mufasa\ \citep{Dave2016} and IllustrisTNG \citep{Nelson2017} is AGN-driven feedback in the low-accretion (jet) mode, the implementation of which has been shown to significantly improve agreement with the observed properties of the $z = 0$ red sequence (e.g. \citealt{Gabor2015}). The typical timescale for quenching due to this process in IllustrisTNG is $\sim1.6$ Gyr, which is consistent with what we observe for case (b). We therefore tentatively associate case (b) with AGN-driven feedback in the low-accretion mode.

Finally, case (c) is associated with some process which leads to a gradual decline in star formation over a period of $\gtrsim$ 3 Gyr. Clearly this  process must be preventative, rather than ejective, however it is the most challenging to associate with a known physical mechanism. As previously noted, this kind of SFH is similar in shape to an exponentially declining or log-normal function. This, and the fact that these objects are observed earliest $(z \simeq 1)$ at the highest stellar masses, suggests a scenario similar to the model of \cite{Gladders2013}. In this model, star formation in individual galaxies naturally dies down in the same way as the cosmic SFR density as the cosmic gas supply gradually dwindles. In the \cite{Gladders2013} model, the rate is determined, in part, by stellar mass, with the most massive galaxies reaching the end-point of their evolution the fastest.

Having tentatively identified the quenching processes at work, it is interesting to consider how their relative importances vary as a function of stellar mass. As was noted in Section \ref{subsubsect:results_how_long_fs_trends}, the distributions of $\tau_\mathrm{quench}$ are remarkably consistent across our bins in stellar mass, which translates into faster physical quenching timescales for more massive galaxies, as we find these to be quenched at higher redshifts on average (Figure \ref{fig:zred_mass_dist}).

This finding is in agreement with previous work (e.g. \citealt{Thomas2010}; \citealt{Pacifici2016}). However the consistency of our distributions of $\tau_\mathrm{quench}$ across our stellar mass bins suggests that faster quenching at higher stellar masses is a further consequence of mass-accelerated evolution (or downsizing), rather than a consequence of a change in quenching mechanism across the range of masses we probe $(M_* > 10^{10}\mathrm{M_\odot})$. The fact that we first begin to see case (c) galaxies (which seem to be associated with a more advanced stage in galaxy evolution) first at the highest masses lends further weight to this theory. The same argument applies to the fact that we observe case (a) galaxies down to lower observed redshifts only at lower masses. These less-massive galaxies appear to be following their more-massive counterparts along the same evolutionary track at a slower pace.

\section{Conclusions}\label{sect:conclusions}

We present the new galaxy spectral model generation framework and fitting tool Bayesian Analysis of Galaxies for Physical Inference and Parameter EStimation, or \bagpipes, which can be used to rapidly generate complex, physically realistic model galaxy spectra and fit these to arbitrary combinations of spectroscopic and photometric data using the \textsc{MultiNest} nested sampling algorithm.

We verify our ability to recover the properties of the SFHs of quenched galaxies by fitting mock observations for a sample of simulated massive quenched galaxies with realistic SFHs, chemical-enrichment histories and dust properties from the \mufasa\ suite of cosmological hydrodynamic simulations.  We investigate thoroughly the effects of our parameterisation and priors on the biases in our derived SFH properties. We find that the use of an exponentially declining SFH model produces a small but significant overestimation of the stellar masses of quenched galaxies, and an overestimation of the redshifts at which their stellar masses were assembled. A significant improvement is observed when fitting a double-power-law SFH model with the priors listed in Table \ref{table:dblplaw_priors}.

We then use \bagpipes\ to perform a detailed analysis of the SFHs of a large sample of 9289 quenched galaxies from the UltraVISTA Survey with stellar masses, $M_* > 10^{10}\ \mathrm{M_\odot}$ across the observed redshift range $0.25 < z_\mathrm{obs} < 3.75$. We make the following observations based on our analysis:

\begin{enumerate}

  \item We observe a clear downsizing trend in our results (Fig. \ref{fig:downsizing}), with the most massive galaxies being on average $\simeq0.5$ Gyr older than those with $M_* \simeq 10^{10}\ \mathrm{M_\odot}$. This appears to remain relatively constant across our whole observed-redshift range.

  \vspace{1.5mm}

  \item A simple analysis of the redshift evolution of the number density of galaxies with very old stellar populations (Section \ref{subsubsect:results_zform_zobs_trend}) suggests that the majority ($61\pm8$ per cent) of galaxies which quench at $z>1.5$ will undergo significant further evolution through rejuvenated star-formation or merger events by $z = 0.5$.

    \vspace{3mm}

  \item The SFHs of our quenched galaxies fall into three distinct classes with different shapes (Fig. \ref{fig:tshutoff}). We refer to these as cases (a), (b) and (c), which roughly correspond to the top, middle and bottom panels of Fig. \ref{fig:example_mufasa_sfh} respectively. We propose that these different cases correspond to different quenching mechanisms (Section \ref{subsubsect:results_how_long_interp}).

  \vspace{3mm}

  \item Galaxies with SFHs of type (a) are common at $z_\mathrm{obs} > 1$. They experience a rapid rise followed by a rapid decline in SFR over a small fraction of the age of the Universe at the redshift of their quenching, and have rapid quenching timescales of $< 1$ Gyr. We tentatively identify these as the result of (possibly merger-triggered) quasar-mode AGN feedback.

  \vspace{3mm}

  \item Galaxies with SFHs of type (b) make up the bulk of our quenched sample across all redshifts and masses. They experience a slower rise and more rapid decline in SFR, with intermediate quenching timescales of $1-2$ Gyr. These SFHs match predictions from simulations for quenching by low-accretion (jet) mode AGN feedback (e.g. \citealt{Dave2017}; \citealt{Nelson2017}).

  \vspace{3mm}

  \item Galaxies with SFHs of type (c) begin to enter our sample at $z_\mathrm{obs} < 1$. Their SFHs rise more rapidly than they fall, and have long quenching timescales of $\gtrsim 3$ Gyr. These are not obviously associated with a specific physical process, and we speculate that they may be the first galaxies in which star formation `naturally' dies down with the decrease of the overall cosmic gas supply.

 \vspace{3mm}

\item In common with other studies, we observe more rapid quenching at higher stellar masses. However, in contrast to some previous studies, we see no clear evidence of a change in the relative importances of different quenching mechanisms with stellar mass (at masses greater than $M_* = 10^{10}$). Our results suggest that this trend is an extension of the well-known mass-accelerated evolution (or downsizing) trend, which appears to affect quenching timescale as well as the redshift of formation.

\end{enumerate}

\noindent To summarise, our results suggest that the red sequence at $M_* > 10^{10}\ \mathrm{M_\odot}$ has been built up since $z \sim 4$ by a number of different quenching processes which make different relative contributions at different epochs. At $z \gtrsim 1$, (potentially merger-triggered) quasar-mode AGN feedback plays a significant role in the rapid but short-lived quenching of galaxies. Throughout cosmic time, AGN feedback in the low-accretion mode is the dominant quenching mechanism, being responsible for the bulk of the build-up of the red sequence. However galaxies which quench by this processes at $z \gtrsim 1$ typically either experience some periods of rejuvenated star formation at later times or merge with younger stellar populations which further reduces the number of extremely early-quenching objects observed at low redshift. Finally, at $z \lesssim 1$, a third mode of `natural' quenching begins to act, shutting down star-formation much more slowly as the cosmic gas supply declines. It is also possible that at lower masses we see rapid quenching due to tidal/ram-pressure stripping of satellite galaxies. This is qualitatively similar to the model of \cite{Schawinski2014}. A joint analysis combining morphology and environment with detailed SFH recovery would provide an additional test of the scheme we propose.

In this work we have considered a large sample of galaxies with high quality photometric data. These data are nevertheless weakly constraining on the SFHs of individual objects (e.g. Fig. \ref{fig:example_mufasa_obj}). We have relied instead upon our large sample size to obtain information about the distribution of the SFH properties of quenched galaxies across bins in stellar mass and observed redshift. This was motivated, in part, by a desire to establish a baseline for how well the SFHs of quenched galaxies could be constrained using photometric observations alone.

Recently, several authors (e.g. \citealt{Pacifici2012}; \citealt{Thomas2017}) have confirmed that significantly better constraints on physical parameters can be obtained by fitting combinations of rest-frame ultraviolet spectroscopy and multi-wavelength photometry. In an upcoming paper we will extend our analysis to fitting spectroscopy from VANDELS (\citealt{McLure2018}; \citealt{Pentericci2018}) with \bagpipes, in order to obtain more detailed information on the properties of individual SFHs, to build up a more detailed understanding of the quenching processes which affect these galaxies.

\section*{Acknowledgements}

The authors all acknowledge the support of the UK Science and Technology Facilities Council. This work is based on data products from observations made with ESO Telescopes at Paranal Observatory under ESO programme ID 179.A2005, and on data products produced by TERAPIX and the Cambridge Astronomy survey Unit on behalf of the UltraVISTA consortium. This work is based in part on observations made with the Spitzer Space Telescope, which is operated by the Jet Propulsion Laboratory, California Institute of Technology under a NASA contract. This research made use of Astropy, a community-developed core Python package for Astronomy \citep{Astropy2013}.

\bibliographystyle{mnras}
\bibliography{carnall2017}

\bsp
\label{lastpage}
\end{document}